\pdfoutput=1
\documentclass[conference]{IEEEtran}

\usepackage[normalem]{ulem}

\usepackage{fancyhdr}
\usepackage[T1]{fontenc}
\usepackage{microtype}
\usepackage{flushend}
\usepackage{ifthen}
\usepackage{textcomp}
\usepackage[vlined,ruled]{algorithm2e}
\usepackage{verbatim}
\usepackage{textgreek}
\usepackage{tikz}
\usetikzlibrary{matrix, plotmarks, chains, positioning, shapes.multipart, calc, shapes, arrows, shapes.arrows}
\usetikzlibrary{calc}
\usepackage{commath}

\usepackage{booktabs}

\usepackage{paralist}
\usepackage{hyperref}
\usepackage{pgfplots}
\usepackage[super]{nth}

\usepackage[shortcuts]{extdash}

\usepackage[nocompress]{cite}

\providecommand{\DontPrintSemicolon}{\dontprintsemicolon}

\newboolean{blindreview}
\setboolean{blindreview}{false}

\newcommand{\void}[1]{}

\newcommand{\todo}[1]{\textbf{\color{red}[Todo: {\small #1}]}}
\renewcommand{\todo}[1]{}

\newcommand{\reftab}[1]{Table~\ref{tab:#1}}
\newcommand{\reffig}[1]{Figure~\ref{fig:#1}}

\newcommand{\refsec}[1]{Section~\ref{sec:#1}}
\newcommand{\refalg}[1]{Algorithm~\ref{alg:#1}}
\newcommand{\refline}[1]{\ref{line:#1}}
\newcommand{\labeltab}[1]{\label{tab:#1}}

\newcommand{\labfig}[1]{\label{fig:#1}}

\newcommand{\labsec}[1]{\label{sec:#1}}
\newcommand{\labalg}[1]{\label{alg:#1}}
\newcommand{\labline}[1]{\label{line:#1}}

\usepackage{xspace}

\newcommand{\negvspacee}{\vspace{0mm}}

\newcommand{\mysubsection}[1]{\negvspacee\subsection{#1}\negvspacee}

\newlength{\tmpTFS}

\newcommand{\microops}{{\textmu}ops\xspace}
\newcommand{\microop}{{\textmu}op\xspace}

\newcommand{\nanoBench}{\emph{nanoBench}\xspace}

\newcommand\copyrighttext{%
\footnotesize \textcopyright\ 2020 IEEE. Personal use of this material is permitted. Permission from IEEE must be obtained for all other uses, in any current or future media, including reprinting/republishing this material for advertising or promotional purposes,creating new collective works, for resale or redistribution to servers or lists, or reuse of any copyrighted component of this work in other works.
}
\def\mycopyrightnotice{
\begin{tikzpicture}[remember picture,overlay]
\node[anchor=south,yshift=10pt] at (current page.south) {\fbox{\parbox{\dimexpr\textwidth-\fboxsep-\fboxrule\relax}{\copyrighttext}}};
\end{tikzpicture}%
}
\makeatletter
\def\ps@IEEEtitlepagestyle{
  \def\@oddfoot{\mycopyrightnotice}
  \def\@evenfoot{}
}

\begin{document}

\title{nanoBench: A Low-Overhead Tool for Running Microbenchmarks on x86 Systems}

\author{\IEEEauthorblockN{Andreas Abel and Jan Reineke}
\IEEEauthorblockA{\textit{Saarland University} \\
\textit{Saarland Informatics Campus}\\
Saarbr\"ucken, Germany \\
\{abel, reineke\}@cs.uni-saarland.de}
}

\maketitle

\thispagestyle{IEEEtitlepagestyle}
\pagestyle{plain}

\begin{abstract}
We present \nanoBench, a tool for evaluating small microbenchmarks using hardware performance counters on Intel and AMD x86 systems.
Most existing tools and libraries are intended to either benchmark entire programs, or program segments in the context of their execution within a larger program. 
In contrast, \nanoBench is specifically designed to evaluate small, isolated pieces of code.
Such code is common in microbenchmark-based hardware analysis techniques.

%der folgenden zwei abstze gehen fr einen abstract schon sehr ins detail
Unlike previous tools, \nanoBench can execute microbenchmarks directly in kernel space.
This allows to benchmark privileged instructions, and it enables more accurate measurements.
The reading of the performance counters is implemented with minimal overhead avoiding functions calls and branches.
%In particular, it does not require the execution of any function calls or branches.
As a consequence, \nanoBench is precise enough to measure individual memory accesses.

%The way \nanoBench runs and evaluates microbenchmarks is highly configurable. %dieser aspekt kommt mir jetzt auf den ersten blick nicht so spektakulr vor, aber vllt. bersehe ich etwas.
%This includes, e.g., the number of times the code of the microbenchmark is unrolled, the number of warm-up runs, and the aggregate function that is applied to the measurement results.
%This eliminates significant engineering efforts that are %typically 
%required when running microbenchmarks with existing tools.

%wrde die "use cases" oder vllt "case studies" strker betonen
We illustrate the utility of \nanoBench at the hand of two case studies.
First, we briefly discuss how \nanoBench has been used to determine the latency, throughput, and port usage of more than 13,000 instruction variants on recent x86 processors.
Second, we show how to generate microbenchmarks to precisely characterize the cache architectures of eleven Intel Core microarchitectures.
This includes the most comprehensive analysis of the employed cache replacement policies to date.

%propose several tools that generate microbenchmarks for determining cache replacement policies, which are typically undocumented for recent microarchitectures.
\end{abstract}

\section{Introduction}

Benchmarking small pieces of code using hardware performance counters is often useful for analyzing the performance of software on a specific microprocessor, as well as for analyzing performance characteristics of the microprocessor itself.

Such microbenchmarks can, for example, be helpful in identifying bottlenecks in loop kernels. 
To this end, modern x86 processors provide many performance events that can be measured, such as cache and TLB hits/misses in different levels of the memory hierarchy, the pressure on execution ports, mispredicted branches, etc.

Low-level aspects of microarchitectures are typically only poorly documented. 
Thus, the only way to obtain detailed information is often through microbenchmarks using hardware performance counters.
This includes, for example, the latency,~throughput, and port usage of individual instructions \cite{Abel19, cqa14,fog17,granlund17,instlatx64}. 
Microbenchmarks have also been used to infer properties of the memory hierarchy~\cite{Saavedra95,Thomborson00, Coleman01,Dongarra04,Yotov05a,Yotov06,Molka09,Babka09,Wong10,Abel12a,Abel13,Abel14,Hassan15,Mei17,Cooper18}.
In addition to that, such benchmarks have been used to identify microarchitectural properties that can lead to security issues, such as Spectre~\cite{kocher19} and Meltdown~\cite{lipp18}.

Often, such microbenchmarks consist of two parts: The main part, and an initialization phase that, for example, sets registers or memory locations to specific values or tries to establish a specific microarchitectural state, for example by flushing the caches.
Ideally, the performance counters should only be active during the main part.

To facilitate the use of hardware performance counters, a number of tools and libraries have been proposed.
Most of the existing tools fall into one of two categories. 
First, there are tools that benchmark entire programs, such as \emph{perf}~\cite{Perf}, or profilers like Intel's VTune Amplifier~\cite{VTune}.
Tools in the second category are intended to benchmark program segments that are executed in the context of a larger program. They usually provide functions to start and stop the performance counters that can be called before and after the code segment of interest. Such tools are, for example, PAPI~\cite{PAPI}, and libpfc~\cite{libpfc}. 
%likwid belongs to both categories; depending on the options that are used, it either benchmarks to whole program, or specific parts.

Tools from both categories are not particularly well suited for microbenchmarks of the kind described above.
For tools from the first category, one obvious reason is that it is not possible to measure only parts of the code.
Another reason is overhead. 
Just running a C program with an empty main function, compiled with a recent version of \emph{gcc}, leads to the execution of more than 500,000 instructions and about 100,000 branches. Moreover, this number varies significantly from one run to another.

Overhead can also be a concern for tools from the second category.
In PAPI, for example, the calls to start and stop the counters involve several hundred memory accesses, more than 150 branches, and for some counters even expensive system calls.
This leads to unpredictable execution times and might, e.g., destroy the cache state that was established in the initialization part of the microbenchmark.
Moreover, these calls will modify general-purpose registers, so it is not possible to set the registers to specific values in the initialization part, and use these values in the main part.

For several reasons, microbenchmarks often need to be run multiple times. 
One reason is the possibility of interference due to interrupts, preemptions or contention on shared resources that are also used by programs on other cores.
Another reason are issues such as cold caches that impact the performance on the first runs.
A third reason is that there are more performance events than there are programmable counters, so the measurements may need to be repeated with different counter configurations.
Also, the code to be benchmarked itself often needs to be repeated several times.\todo{why?}
This is typically done by executing it in a loop or by unrolling it multiple times, or by a combination of both.
All of this leads to a significant engineering effort that needs to be repeated over and over again.\looseness=-1
\todo{Diesen Punkt finde ich eher schwach. ist es so schwierig ein paar Schleifen zu programmieren? (könnte ein Kritiker sagen...)}

In this paper, we present \nanoBench, an open-source tool that was developed to make it very easy to execute microbenchmarks consisting of small, independent pieces of machine code on recent x86 CPUs. 
\nanoBench is available on GitHub\footnote{\url{https://github.com/andreas-abel/nanoBench}}.

There are two variants of the tool: A user-space implementation and a kernel-space version. 
The kernel-space version makes it possible to directly benchmark privileged instructions, in contrast to any previous tool we are aware of.
Furthermore, it allows for more accurate measurements than existing tools by disabling interrupts and preemptions. 
The tool is precise enough to measure, e.g., whether individual memory accesses result in cache hits or misses.

Microbenchmarks may use and modify any general-purpose and vector registers, including the stack pointer.
After executing the microbenchmark, \nanoBench automatically resets them to their previous values.
The loop and unroll counts, as well as the number of repetitions and the aggregate function to be applied to the measurement results, can be specified via parameters.

To demonstrate the usefulness of our tool, we present two case studies.

First, we discuss how \nanoBench is used to obtain the latency, throughput, and port usage data for more than 13,000 instruction variants that is available at \href{https://www.uops.info/}{\emph{uops.info}}.

For the second case study, we develop a set of tools that generate microbenchmarks for analyzing caches. 
These microbenchmarks are then evaluated using \nanoBench.
We focus, in particular, on cache replacement policies, which are typically undocumented.
We apply our tools to eleven different Intel microarchitectures, and provide detailed models of their replacement policies, including several previously undocumented variants.

\subsubsection*{Summary of contributions}
Our primary contribution is \nanoBench, a low-overhead tool for evaluating small microbenchmarks in a highly accurate and precise way.
As a secondary contribution, we use nanoBench to obtain comprehensive characterizations of the cache replacement policies employed in recent Intel microarchitectures.
A third contribution is the extension of the work described in~\cite{Abel19} to seven additional microarchitectures
(including two from AMD), and more than 10,000 additional instruction variants.

\section{Background}\labsec{background}
Recent Intel and AMD processors are equipped with different types of performance counters.
All of these counters can be read using the \emph{RDMSR}\footnote{``Read from model specific register''} instruction; many of them can also be read using the \emph{RDPMC}\footnote{``Read performance-monitoring counters''} instruction.
The \emph{RDMSR} instruction is a privileged instruction, and can thus only be used in kernel space.
The \emph{RDPMC} instruction, on the other hand, is faster than the \emph{RDMSR} instruction, and it can be directly accessed in user space if a specific flag in a control register is set.
%documented in the manuals
\subsection{Core Performance Counters}
Each logical core has a private performance monitoring unit with multiple performance counters.
\medskip

\subsubsection{Fixed-Function Performance Counters}
Recent Intel CPUs have three fixed-function performance counters that can be read with the \emph{RDPMC} instruction. 
They count the number of retired instructions, the number of core cycles, and the number of reference cycles.
\todo{if space permits, explain difference between core and reference cycles}

In addition to that, there are two fixed-function counters that are available both on recent Intel CPUs, as well as on AMD family 17h CPUs: the \emph{APERF} counter, which counts core clock cycles, and the \emph{MPERF} counter, which counts reference cycles.
These two counters can only be accessed with the \emph{RDMSR} instruction, and are thus only available in kernel space.
\medskip

\subsubsection{Programmable Performance Counters}
Recent Intel CPUs have between two and eight, and AMD family 17h CPUs have six programmable performance counters.
They can be programmed with a large number of different performance events (more than 200 on some CPUs), such as the number of \microops that use a specific port, the number of cache misses in different levels of the memory hierarchy, the number of mispredicted branches, etc.
These counters can be read with the \emph{RDPMC} instruction.
\subsection{Uncore/L3 Performance Counters}\labsec{uncoreCounters}
In addition to the per-core performance counters described above, recent processors also have a number of global performance counters that can, in particular, count events related to the shared L3 caches.
On Intel CPUs, these counters can only be read in kernel space.

\section{\nanoBench Features}\labsec{features}

In this section, we will first give a high-level overview by looking at a simple example that shows how \nanoBench can be used.
We will then describe various features of \nanoBench in more detail.

\subsection{Example}\labsec{example}
The following example shows how \nanoBench can be used to measure the latency of the L1 data cache on a Skylake-based system.
\begin{verbatim}
./nanoBench.sh -asm "mov R14, [R14]"
               -asm_init "mov [R14], R14"                
               -config cfg_Skylake.txt
\end{verbatim}

The tool will first execute the instruction $$\verb|mov [R14], R14|$$ which copies the value of register \verb|R14| to the memory location that \verb|R14| points to.
\nanoBench always initializes \verb|R14| (and a number of other registers) to point into a dedicated memory area that can be freely modified by microbenchmarks; this is described in more detail in~\refsec{AccMem}.

\nanoBench then starts the performance counters, and executes the instruction $$\verb|mov R14, [R14]|$$ multiple times. 
The number of repetitions can be controlled via parameters; for more information see \refsec{Loop}.
The instruction loads the value at the address in \verb|R14| into \verb|R14|. Thus, the execution time of this instruction corresponds to the L1 data cache latency.
Afterwards, \nanoBench stops the performance counters.

The entire benchmark is then repeated multiple times to obtain stable results.

The output of \nanoBench will be similar to the following:
\begin{verbatim}
Instructions retired: 1.00
Core cycles: 4.00
Reference cycles: 3.52
UOPS_ISSUED.ANY: 1.00
UOPS_DISPATCHED_PORT.PORT_0: 0.00
UOPS_DISPATCHED_PORT.PORT_1: 0.00
UOPS_DISPATCHED_PORT.PORT_2: 0.50
UOPS_DISPATCHED_PORT.PORT_3: 0.50
MEM_LOAD_RETIRED.L1_HIT: 1.00
MEM_LOAD_RETIRED.L1_MISS: 0.00
\end{verbatim}

The first three lines show the result of the fixed-function performance counters.
The remaining lines correspond to the performance events specified in the \emph{cfg\_Skylake.txt} configuration file that was supplied as a parameter in the \nanoBench call shown above; details on the configuration file are described in \refsec{configFile}.

From the results, we can conclude that the L1 data cache latency is 4 cycles. 
This agrees with the documentation in Intel's optimization manual~\cite{intelOptManual19}.

\subsection{Generated Code}
To execute a microbenchmark, \nanoBench first generates code for a function similar to the pseudocode shown in \refalg{genCode}.
In line~\refline{saveRegs}, the generated code first saves the current values of the registers to the memory, and initializes certain registers to point to specific memory locations (see \refsec{AccMem}).
Then, the initialization part of the microbenchmark is executed (line~\refline{codeInit}).
In the next line (line~\refline{perf1}), the performance counters are read.
Unless the noMem option (see \refsec{noMemMode}) is used, this step does not modify the values in any general-purpose or vector registers that were set by the initialization code (technically, it does modify certain registers temporarily, but it resets them to their previous value before the next line is executed).
Lines \refline{for1} to \refline{codeN} contain the code for the main part of the microbenchmark.
The code is unrolled multiple times (this can be configured via a parameter, see \refsec{Loop}).
If the parameter \emph{loopCount} is larger than $0$, the code for a for-loop is inserted in line~\refline{for1}; in this case, the code of the microbenchmark must not modify register \verb|R15|, which is used to store the loop counter.
%Otherwise, this line is omitted, but the code is still unrolled \emph{unrollCount} many times.
Afterwards, the performance counters are read a second time (line~\refline{perf2}), and in line~\refline{restoreRegs}, the registers are restored to the values that were saved in line~\refline{saveRegs}.
Finally, the difference between the two performance counter values, divided by the number of repetitions, is returned.

\LinesNumbered
\newcommand{\assign}{\leftarrow}
\begin{algorithm}[t]
\labalg{genCode}
\caption{Generated Code for a Microbenchmark}
\DontPrintSemicolon
%\SetAlgoVlined
%saveState() \label{saveState}\\

\SetKwProg{Fn}{Function}{}{}
\Fn{generatedCode()}{    
    saveRegs \labline{saveRegs}\;
    codeInit \labline{codeInit}\;
    \textit{m1} $\assign$ readPerfCtrs \tcp*{stores results in memory, does not modify registers} \labline{perf1}
    \For(\tcp*[f]{this line is omitted if loopCount=0}){$j \assign 0$ \textbf{\upshape to} \textit{loopCount}}{ \labline{for1}
        code \tcp{copy \#1} \labline{code1}
        code \tcp{copy \#2}
        $\vdots$ \;
        code \tcp{copy \#localUnrollCount} \labline{codeN}
    }
    \textit{m2} $\assign$ readPerfCtrs \labline{perf2}\;
    restoreRegs \labline{restoreRegs}\;
    \textit{r} $\assign$ \textit{(m2-m1)/(max(1,loopCount)*localUnrollCount)}\;
    \Return{r}
}

\end{algorithm}

\begin{algorithm}[t]
\labalg{run}
\caption{Running a Microbenchmark}
\DontPrintSemicolon
\SetKwProg{Fn}{Function}{}{}
\Fn{run(code)}{
    \For{$i \assign -\textit{warmUpCount}\,\ \textbf{\upshape to}\ \textit{nMeasurements}$}{
        $m \assign \textit{code()}$\;
        \If(\tcp*[f]{ignore warm-up runs}){
           $i \geq 0$}{\textit{measurements}[\textit{i}] $\assign$ \textit{m}
        }
    } 
    \tcp{apply aggregate function}
    \Return{agg(measurements)} 
}
\end{algorithm}

\subsection{Running the Generated Code}
\refalg{run} shows how the generated code is run.
The code is run a configurable number of times.
At the end, an aggregate function is applied to the measurement results, which can be either the minimum, the median, or the arithmetic mean (excluding the top and bottom 20\% of the values).
A~configurable number of runs in the beginning can be excluded from the result; this is described in more detail in \refsec{warmup}.

By default, \nanoBench generates and runs two versions of the code: the first one with \emph{localUnrollCount} set to the specified \emph{unrollCount}, and the second time with \emph{localUnrollCount} set to two times the specified \emph{unrollCount}.
The reported result is the difference between the two runs.
This removes the overhead of the measurement instructions from the result, as well as anomalies that might be caused by the serialization instructions that are needed before and after reading the performance counters (see also~\refsec{serialization}).

\nanoBench also provides an option that uses a \emph{localUnrollCount} of 0 for one of the runs instead (i.e., there are no instructions between line~\refline{perf1} and line~\refline{perf2} in this case).

\subsection{Kernel/User Mode}
\nanoBench is available in two versions: A user-space and a kernel-space version.

The kernel-space version has several advantages over the user-space version:
\begin{itemize}
\item It makes it possible to benchmark privileged instructions.
\item It can allow for more accurate measurement results as it disables interrupts and preemptions during measurements.
\item It can use several performance counters that are not accessible from user space, like the uncore counters on Intel CPUs, or the APERF and MPERF counters on AMD.
\item It can allocate physically-contiguous memory. See also \refsec{AccMem}.
\end{itemize}

On the other hand, executing microbenchmarks in kernel space can lead to potential data loss and security problems, if the microbenchmarks contain bugs.
It is thus recommended to use the kernel-space version only on dedicated test machines.

\subsection{Interface}\labsec{interface}
We provide a unified interface to the user-space and the kernel-space version in the form of two shell scripts, \verb|nanoBench.sh| and \verb|kernel-nanoBench.sh|, that have mostly the same command-line options.

In addition to that, we also provide a Python interface for the kernel-space version.
This interface is used for the case studies in Sections \ref{sec:useCaseInstr} and \ref{sec:useCaseCaches}.

With all interfaces, the code of the microbenchmarks can be specified either as an assembler code sequence in Intel syntax (like in the example in \refsec{example}), or by the name of a binary file containing x86 machine code.

\subsection{Loops vs. Unrolling}\labsec{Loop}
For microbenchmarks that have code that needs be repeated several times to obtain meaningful results, there is a trade-off between unrolling the code (i.e., creating multiple copies of it), and executing the code in a loop.

Using a loop has the advantage of keeping the code size small, so that it will fit into the cache.
On the other hand, the loop introduces an additional overhead, which can be significant if the body of the loop is small.

Whether unrolling or a loop should be used, depends on the particular benchmark.
For benchmarks that measure, e.g., the number of data cache misses, a loop is the better choice, as it does not introduce any overhead in terms of memory accesses.
On the other, for a benchmark that measures the port usage of an instruction, using only unrolling is better, as otherwise, the \microops of the loop code compete for ports with the \microops of the benchmark.

For some benchmarks, a combination of both a loop and unrolling yields the best results.

\nanoBench provides two parameters, \emph{loopCount} and \emph{unrollCount}, that control the number of loop iterations, and how often the code is unrolled.

\subsection{Accessing Memory}\labsec{AccMem}
\nanoBench initializes the registers \verb|RSP| (i.e., the stack pointer), \verb|RBP| (i.e., the base pointer), \verb|RDI|, \verb|RSI|, and \verb|R14| to point into dedicated memory areas (of 1 MB each) that can be freely modified by the microbenchmarks. \todo{it might be good to elaborate by giving examples where this is required}

Furthermore, for microbenchmarks needing a larger memory area, like benchmarks for determining cache parameters, the kernel-version of \nanoBench provides an option for reserving a physically-contiguous memory area of a specific size that register~\verb|R14| points to (see also~\refsec{allocMem}).

\subsection{Warm-Up Runs}\labsec{warmup}
\nanoBench provides the option of performing a configurable number of initial benchmark runs that are excluded from the results.
This can, for example, be useful to make sure that the code and other accessed memory locations are in the cache.
It can also be used to train the branch predictor to reduce the number of mispredicted branches.
Furthermore, there are some instructions that require a warm-up period after having not been used for a while before they can execute at full speed again, like \emph{AVX2} instructions on some microarchitectures.

\subsection{noMem Mode}\labsec{noMemMode}
By default, the code to read the performance counters writes the results to the memory.
After a warm-up run, this memory location is usually in the cache, and thus, the time for these memory operations is constant.

However, for microbenchmarks that contain many memory accesses to different addresses that map to the same cache set, writing the performance counter results to the memory can be problematic.
One reason for this is that the memory accesses in line~\refline{perf1} may change a cache state that was established by the initialization part of the benchmark.
Another reason is that the microbenchmark code may evict the block that stores the performance counter results, which would lead to additional cache misses.

To avoid these problems, \nanoBench has a special mode that stores all performance counter measurements in registers instead of in memory.
If this mode is used, certain general-purpose registers must not be modified by the microbenchmark. 
%precise enough to measure individual memory accesses

Moreover, if this mode is used, \nanoBench also provides a feature to temporarily pause performance counting.
This feature can be used by including special magic byte sequences in the microbenchmark code for stopping and resuming the performance counters.
Using this feature incurs a certain timing overhead, so it is in particular useful for benchmarks that do not measure the time but, e.g., the number of cache hits or misses.

\subsection{Performance Counter Configurations}\labsec{configFile}
The performance events to be measured are specified in a configuration file.
The file uses a simple syntax to define the events.
Unlike in some previous tools, like libpfc~\cite{libpfc}, the events are not hard-coded, which makes it easy to adapt \nanoBench to future CPUs, as only a new configuration file has to be created.

If the configuration file contains more events than there are programmable performance counters, the benchmarks are automatically executed multiple times with different counter configurations.

We provide configuration files with all events for all recent Intel microarchitectures, and the AMD Zen microarchitecture.

\subsection{Execution Time of nanoBench}
Evaluating microbenchmarks with \nanoBench is very fast.
As an example, we consider a benchmark consisting of a single \emph{NOP} instruction, that is run with \emph{unrollCount}~=~100, \emph{loopCount}~=~0, \emph{nMeasurements}~=~10, and a configuration file with four events.
On an Intel Core i7-8700K, running \nanoBench with these parameters takes about $15 ms$ for the kernel version (assuming that the kernel module is already loaded), and about $50 ms$ for the user-space version. 

\subsection{Supported Platforms}
We have successfully used \nanoBench on processors from most generations of Intel's Core microarchitecture, and with AMD Ryzen CPUs. 
All experiments were performed under Ubuntu 18.04, but \nanoBench should be compatible with any Linux distribution that uses a recent kernel version.

\section{\nanoBench Implementation}
In this section, we describe several aspects of our implementation.
%configuring perf counters in user space

\subsection{Accurate Performance Counter Measurements}
\subsubsection{Serializing Instruction Execution}\labsec{serialization}
As described in \refsec{background}, performance counters can be read with the \emph{RDPMC} or the \emph{RDMSR} instruction.
These instructions are not serializing instructions.
Thus, due to out-of-order execution, they may be reordered with earlier or later instructions by the processor.
For obtaining meaningful measurement results, it is therefore important to add instructions that serialize the instruction stream both before and after any instructions that read performance counters.

Previous approaches (e.g., \cite{fogTest}) often use the \emph{CPUID} instruction for that purpose.
However, for benchmarking short code segments, this is problematic.
One reason for this is that the \emph{CPUID} instruction has a variable latency and \microop count. 
Paoloni~\cite{paoloni2010benchmark} observed that the execution time of the \emph{CPUID} can differ by hundreds of cycles from run to run.
The variable \microop count can be eliminated by setting the register \verb|RAX| to a fixed value before each execution of  the \emph{CPUID} instruction; this also reduces the variance in the execution time, but does not fully eliminate it.
Moreover, for an instruction sequence of the form \verb|A; CPUID; B|, the serialization property of the \emph{CPUID} instruction only guarantees that all \microops of \verb|A| have completed before \verb|B| is fetched and executed.
It does not guarantee that all \microops of \verb|A| have completed before the first \microop of the \emph{CPUID} instruction is executed, and it does also not guarantee that all \microops of the \emph{CPUID} have completed before the first \microop of \verb|B| is executed.

We propose to use the \emph{LFENCE} instruction instead. 
This instruction is not fully serializing: it does not guarantee that earlier stores have become globally visible, and subsequent instructions may be fetched from memory before \emph{LFENCE} completes.
However, on Intel CPUs it does guarantee that ``\emph{LFENCE} does not execute until all prior instructions have completed locally, and no later instruction begins execution until \emph{LFENCE} completes.''~\cite{intelDevManual}. 
For our purposes, this is sufficient, and the guarantee is even somewhat stronger than that for the \emph{CPUID} instruction, as it also orders the \emph{LFENCE} instruction itself with respect to the preceding and succeeding instructions.
On AMD CPUs, the \emph{LFENCE} provides similar guarantees if Spectre mitigations are enabled.

Using the \emph{LFENCE} instruction for measurements of short durations was also recently recommended by McCalpin~\cite{McCalpin18}.
\medskip

\subsubsection{Reducing Interference}\labsec{reducingInterference}
In the kernel-space version, we disable preemptions and hard interrupts during measurements, as they can perturb the measurement results~\cite{weaver2008can, weaver2013non}.
This is not possible for the user-space version; however, we do pin the process to a specific CPU in this case to avoid the cost of process switches between CPUs.

Furthermore, for obtaining unperturbed measurement results, we recommend disabling hyperthreading.
When using performance counters for resources shared by multiple cores, such as L3 caches, we furthermore recommend disabling all cores that share these resources.
We provide shell scripts for this in our repository.

For microbenchmarks that measure properties of caches, such as the benchmarks described in \refsec{useCaseCaches}, it can be helpful to disable cache prefetching. 
On Intel CPUs, this can be achieved by setting specific bits in a model-specific register (MSR). 
Details on how to do this are available in the documentation of \nanoBench.

\subsection{Generating Code}\labsec{generatingCode}
As described in \refsec{features}, \nanoBench runs microbenchmarks by generating a function that contains the code of the microbenchmark, as well as setup and measurement instructions.
This is implemented by first allocating a large enough memory area, and marking it as executable.
Then, the corresponding machine code is written to this memory area, including \emph{unrollCount} many copies of the code of the microbenchmark.
If this code contains the magic byte sequences for pausing performance counting as described in \refsec{noMemMode}, they are replaced by corresponding machine code for reading performance counters.

Generating the code for executing the microbenchmarks at runtime in this way makes it possible to access the performance counters without having to execute any function calls or branches.

\subsection{Kernel Module}
The kernel-space version of \nanoBench is implemented as a kernel module.
While the module is loaded, it provides a set of virtual files that are used to configure and run microbenchmarks.
For example, setting the loop count, or the code of the microbenchmark is done by writing the corresponding values to specific files under \verb|/sys/nb/|.
Reading the file \verb|/proc/nanoBench| generates the code for running the benchmark (as described in \refsec{generatingCode}), runs the benchmark (possibly multiple times, depending on the configuration), and returns the result of the benchmark.

Note that it is usually not necessary to access these virtual files directly, as we provide convenient interfaces that perform these accesses automatically (see \refsec{interface}).

\subsection{Allocating Physically-Contiguous Memory}\labsec{allocMem}
In Linux kernel code, the \emph{kmalloc} function can be used to allocate physically-contiguous memory.
With recent kernel versions, this is limited to at most 4 MB.

Some of the microbenchmarks for determining properties of the L3 caches that we describe in \refsec{useCaseCaches} require larger memory areas.
We are not aware of a way to directly allocate larger physically-contiguous memory areas.
However, we noticed that in many cases, subsequent calls to \emph{kmalloc} yield adjacent memory areas.
This is, in particular, the case if the system was rebooted recently. 
Moreover, the corresponding virtual addresses are also adjacent.

Based on this observation, we implemented a greedy algorithm that tries to find a physically-contiguous memory area of the requested size by performing multiple calls to \emph{kmalloc}.
If this does not succeed, the tool proposes a reboot.
Note that allocating memory is only necessary once when the kernel module is loaded, and not before each microbenchmark run.

\section{Case Study I: Instruction Latencies, Throughputs, and Port Usages}\labsec{useCaseInstr}

We developed an approach to automatically generate assembler code for microbenchmarks that measure the latencies, throughputs,
and port usages of x86 instructions on Intel and AMD microarchitectures, which are often undocumented.
For the latency, our approach considers dependencies between different pairs of input and output
operands; we take into account explicit and implicit dependencies, such as, e.g., dependencies on status flags.

The generated microbenchmarks are then evaluated using \nanoBench.
Of particular use is \nanoBench's ability to benchmark privileged instructions, the ability to unroll the code multiple times, and the support for microbenchmarks to have an initialization sequence that is not part of the performance measurement. 
Such an initialization sequence is often needed to, e.g., set registers or memory locations to specific values, for example, valid floating numbers if the microbenchmark uses floating point instructions.

More details on our approach have been published in~\cite{Abel19}.
We have since extended our approach to also support AVX-512 instructions; with this extension, the tool is now able to automatically obtain latency, throughput, and port usage data for more than 13,000 instruction variants.
We have applied our implementation to additional microarchitectures, including Intel's Cannon Lake and Ice Lake microarchitectures, and AMD's Zen+ and Zen 2 microarchitectures.
Our results are available at \href{https://www.uops.info/}{\emph{uops.info}} both in the form of a human-readable, interactive HTML table, and as a machine-readable XML file.\looseness=-1

\section{Case Study II: Caches}\labsec{useCaseCaches}
For our second case study, we develop a set of tools that generate microbenchmarks for analyzing caches.
We focus, in particular, on cache replacement policies, which are typically undocumented for recent microarchitectures.

%\todo{needs some introduction: what are the goals? what is unknown? ...}

\mysubsection{Background on Cache Organization}

\newcommand{\blockf}{\mathrm{block}}
\newcommand{\ways}{Way}
\newcommand{\indices}{Index}
\newcommand{\setstate}{\mathrm{SetState}}
\newcommand{\logicalsetstate}{\mathrm{LogicalSetState}}
\newcommand{\cachestate}{\mathrm{CacheState}}
\newcommand{\policystate}{\mathrm{PolState}}
\newcommand{\indexf}{\mathrm{index}}
\newcommand{\upf}{\mathrm{up}}
\newcommand{\hitf}{\mathrm{hit}}
\newcommand{\replupf}{\mathrm{up}}
\newcommand{\evictf}{\mathrm{evict}}
\newcommand{\myvec}[1]{{\langle}#1\rangle}
\newcommand{\associativity}{A}
\newcommand{\numberofsets}{N}
\newcommand{\blocksize}{B}	
\newcommand{\capacity}{C}
\newcommand{\waysize}{W}
\newcommand{\policy}{P}
\newcommand{\LRUPLRU}{\emph{LRU\textsubscript{3}PLRU\textsubscript{4}}\xspace}

To profit from spatial locality and to reduce management overhead, main memory is logically partitioned into a set of \emph{memory blocks} of a specific size (typically 64 Bytes).
Blocks are cached as a whole in cache lines of the same size.
%Usually, the block size is a power of two.
%This way, the block number is determined by the most significant bits of a memory address.
%, more generally: $\blockf_{\blocksize}(address) = \lfloor{address/\blocksize}\rfloor$.

When accessing a memory block, the cache logic has to determine whether the block is stored in the cache (``cache hit'') or not (``cache miss'').
To enable an efficient lookup, each block can only be stored in a small number of cache lines.
For this purpose, caches are partitioned into $\numberofsets$ equally-sized \emph{cache sets}.
The size of a cache set is called the \emph{associativity}~\associativity\ of the cache.
A cache with associativity~\associativity\ is often called \associativity-\emph{way} set-associative.
It consists of \associativity~\emph{ways}, each of which consists of one cache line in each cache set.
%In the context of a cache set, the term \emph{way} thus refers to a single cache line.
%Usually, also the number of cache sets $\numberofsets$ is a power of two such that the set number, also called \emph{index}, is determined by the least significant bits of the block number.
%More generally: $\indexf_{\blocksize, \numberofsets}(address) = \blockf_{\blocksize}(address)\mod \numberofsets.$
%The remaining bits of an address are known as the \emph{tag}: $\tagf_{\blocksize, \numberofsets}(address) = {\lfloor\blockf_{\blocksize}(address)/\numberofsets\rfloor}$.
%To decide whether and where a block is cached within a set, tags are stored along with the data.

In Intel microarchitectures, starting with Sandy Bridge, the last-level cache is divided into multiple slices.
Each of the slices is organized as described above.
The slices are managed by so called C-Boxes, which provide the interface between the core and the last-level cache, and which are responsible for maintaining cache coherence.
Usually, there is one C-Box per physical core.
The first microarchitectures that used sliced L3 caches (Sandy Bridge, Ivy Bridge, Haswell) had one slice per C-Box~\cite{hund13, irazoqui15, liu15, Maurice15, yarom15, inci16, kayaalp16}.
Skylake and more recent microarchitectures can have multiple slices per C-Box~\cite{Disselkoen17, Farshin19}.
Each C-Box has several performance counters that can, e.g., count the number of lookup events for the corresponding part of the last-level cache.
These counters belong to the class of \emph{uncore performance counters} (see \refsec{uncoreCounters}).

An undocumented hash function is used for mapping physical addresses to cache slices.
This hash function has been reverse-engineered for Sandy Bridge, Ivy Bridge, and Haswell CPUs~\cite{hund13, irazoqui15, Maurice15, yarom15, inci16, kayaalp16}.

\subsection{Background on Replacement Policies}
Since the number of memory blocks that map to a set is usually far greater than the associativity of the cache, a \emph{replacement policy} must decide which memory block to replace upon a cache miss.
\medskip
%Most replacement policies try to exploit temporal locality and base their decisions on the history of memory accesses.
%Usually, cache sets are treated independently of each other such that accesses to one set do not influence replacement decisions in other sets.

\subsubsection{Permutation-Based Policies}

Many commonly used policies can be modeled as so called \emph{permutation policies}. These policies have in common that  
\begin{itemize}
\item they maintain a total order of the elements in the cache,
\item upon a cache hit, the order is updated; the new order only depends on the position of the accessed element in the order, and
\item upon a cache miss, the smallest element in the order is replaced.
\end{itemize}
Permutation policies can thus be fully specified by $\associativity+1$ many permutations (one for each position in which a hit can occur, and one permutation for a miss). 

Among the policies that can be modeled as \emph{permutation policies} are, for example, FIFO, LRU, and tree-based pseudo-LRU (PLRU).
PLRU is an approximation to LRU that maintains a binary search tree for each cache set. Upon a cache miss, the element that the tree bits currently point to is replaced. After each access to an element, all the bits on the path from the root of the tree to the leaf that corresponds to the accessed element are set to point away from this path.

Permutation policies were introduced by \cite{Abel13}, along with an efficient algorithm for inferring them automatically.
\medskip

%\subsubsection{Bit-based policies}\labsec{bitPol}
%\todo{ist ``bit-based'' etablierte Terminologie?}

%A more general class of policies that subsumes permutation policies are bit-based replacement policies.
%These policies store one or more status bits per cache line.
%Upon an access, the status bits of all lines may be updated.
%The element to be replaced upon a miss is determined based on the status bits and/or the physical location in the cache.
\subsubsection{MRU/QLRU}\labsec{bitPol}
However, not all popular policies can be modeled as permutation policies.

One example is the MRU policy~\cite{reineke07}.\todo{let's call this NRU and mention MRU as a name used elsewhere. it's the better name... - NRU, as described in jaleel10, is actually a slightly different policy}
%\todo{i'd rather cite the earlier work discussing this policy, e.g. \cite{al04}}
This policy stores one status bit for each cache line. 
Upon an access to a line, the corresponding bit is set to zero; if it was the last bit that was set to one before, the bits for all other lines are set to one.
Upon a cache miss, the leftmost element whose bit is set to one gets replaced.
This policy is sometimes also called bit-PLRU~\cite{pan15} or PLRUm~\cite{al04}.
A variant of this replacement policy that only checks upon a cache miss whether there is still a line whose status bit is one, is called not-recently-used (NRU)~\cite{jaleel10}. 

A generalization of this policy that uses two status bits per cache line is called Quad-Age LRU (QLRU)~\cite{jahagirdar12, briongos19}, or ``2-bit Re-reference Interval Prediction'' (RRIP)~\cite{jaleel10}.
The two bits are supposed to represent the age of a block. 

During our experiments, we found out that some recent Intel CPUs use variants of this policy that were not described in the literature so far.
In particular, the variants differ from each other in the \emph{hit promotion policy}, in the \emph{insertion age}, in the location in the cache where a block is inserted upon a miss, in how the bits are updated if there is no more block with age $3$, and in whether this update occurs only on a miss, or also on a hit.
In the following, we will describe these parameters in detail, and we propose a naming scheme for referring to the different variants.

The \emph{hit promotion policy} describes how the age of a block is updated upon a hit.
We assume that the age is always reduced, unless it is already $0$.
Thus, the hit promotion policy can be modeled by one of the following functions.
Let $x \in \{0,1,2\}$, and $y \in \{0,1\}$.
\begin{equation*}
    Hxy(a) := \begin{cases}
               x,     & \text{if } a = 3\\
               y,     & \text{if } a = 2\\
               0,	 & \text{otherwise}
           \end{cases}
\end{equation*}

The \emph{insertion age} is the age that will be assigned to a block upon a miss.
For $x \in \{0,1,2,3\}$, we will use $Mx$ to denote that the insertion age is $x$.

Furthermore, we will use $\textit{MR}_px$ to denote a policy that inserts new blocks with age $x$ with probability $\frac{1}{p}$, and with age $3$ otherwise.

Note that the insertion age might be different if blocks are brought into the cache by prefetching. 
We currently do not consider this scenario.

We consider the following three variants as to where a block will be inserted upon a miss.
\begin{itemize}
\item\emph{R0:} If the cache is not yet full (after executing the \emph{WBINVD} instruction), insert the new block in the leftmost empty location. Otherwise, replace the block in the leftmost location whose status bits are $3$.
If there is no such block, the behavior is undefined.
\item\emph{R1:} Like \emph{R0}, but if there is no location whose status bits are $3$, always replace the leftmost block, independently of its status bits.
\item\emph{R2:} Like \emph{R0}, but insert blocks in the rightmost empty location if the cache is not yet full.
\end{itemize}

If after an access, there is no more block whose age is $3$, the status bits of potentially all blocks will be updated. 
Let $i$ denote the location of the block that was accessed.
Let $age(b)$ be the current age of block $b$, and $age'(b)$ the new age (after the update).
Let $M$ be the maximum (current) age of any block.
We consider the following variants for $age'$:
\begin{itemize}
\item\emph{U0:} $age'(b) := age(b) + (3 - M)$
\item\emph{U1:} $
    age'(b) := \begin{cases}
               age(b),     			& \text{if } b = i\\
               age(b) + (3 - M), 	& \text{otherwise}
           \end{cases}
$
\item\emph{U2:} $
    age'(b) := age(b) + 1
$
\item\emph{U3:} $
    age'(b) := \begin{cases}
               age(b),     			& \text{if } b = i\\
               age(b) + 1, 			& \text{otherwise}
           \end{cases}
$
\end{itemize}

We will use a name of the form \emph{QLRU\_H11\_M1\_R1\_U2} to refer to the corresponding variant.

Some variants do not check after each access whether there is still a block with age $3$, as described above, but only upon a miss, before selecting the block to replace.
We will refer to such variants by adding the suffix \verb|UMO| (``update on miss only'') to the name.

Note that not all combinations are possible.
For example, \emph{R0} cannot be combined with \emph{U2} or \emph{U3}, as it always requires at least one block with age $3$.
Also, some combinations are observationally equivalent; this is, e.g., the case for \emph{R0} and \emph{R1} in combination with \emph{U0}.

The \emph{2-bit SRRIP-HP} policy proposed by~\cite{jaleel10} would be named \emph{QLRU\_H00\_M2\_R0\_U0\_UMO} according to our naming scheme.
The corresponding ``bimodal RRIP'' (BRRIP) policy from the same paper would be named \emph{QLRU\_H00\_MR\textsubscript{p}2\_R0\_U0\_UMO}.

\subsubsection{Adaptive Policies}
Some caches use adaptive replacement policies that can dynamically switch between two different policies.
This can be implemented via \emph{set dueling}~\cite{qureshi07, jaleel10, wong13}: A number of sets are dedicated to each policy, and the remaining sets are \emph{follower sets} that use the policy that is currently performing better.

\newcommand{\wbinvd}{\langle\textrm{{\upshape}wbinvd}\rangle}
\subsection{Cache-Characterization Tools}\labsec{cacheTools}
Based on \nanoBench, we have developed a set of tools for analyzing undocumented properties of caches.

The first tool, \emph{cacheSeq}, can be used to measure how many cache hits and misses executing an access sequence (i.e., a sequence of blocks that map to the same cache sets) generates.
To this end, \emph{cacheSeq} automatically generates a suitable microbenchmark that is then evaluated using the kernel-space version of \nanoBench.

Access sequences can be specified using strings of the following form:
$$\textrm{``}A\ \wbinvd\ B_0\ B_1\ B_2\ B_3\ B_0?\ B_1!\ X\ A?\textrm{''}$$
Elements of the sequence that end with a ``$?$'' will be included in the performance counter measurements.
The other elements will be accessed, but the number of hits and misses that they generate will not be recorded; this is implemented using the feature described in~\refsec{noMemMode}, that makes it possible to temporarily pause performance counting.
Elements that end with a ``$!$'' will be flushed (using the \emph{CLFLUSH} instruction) instead of being accessed.
``$\wbinvd$'' means that the \emph{WBINVD}\footnote{``Write back and invalidate cache''} instruction will be executed at the corresponding location in the access sequence. This instruction, which is a privileged instruction, flushes all caches.

The following parameters can be specified via command-line options:
\begin{itemize}
\item The cache level, in which the sequence should be accessed.
\item The cache sets, in which the sequence should be accessed. This can be a list or a range of sets.
\item The C-Box, in which the sequence should be accessed. If there are multiple slices per C-Box, all accesses will be to the same slice.
\item The sequence can be executed a configurable number of times in a loop. Additionally, it is possible to specify an initialization sequence that is executed once in the beginning.
\item Between every two accesses to the same set in a lower-level cache, \emph{cacheSeq} can automatically add a sufficient number of accesses to the higher-level caches (that map to different sets and/or slices in the lower-level cache) to make sure that the corresponding lines are evicted from the higher-level cache and the access actually reaches the lower-level cache. These additional accesses are excluded from the performance counter measurements.
\end{itemize}

The syntax for specifying access sequences was inspired by an early version of Vila et al.'s \emph{MemBlockLang} (MBL) language~\cite{Vila19b}.

The following tools are all based on \emph{cacheSeq}.
\medskip

\subsubsection{Replacement Policies}\labsec{toolReplPol}
We implemented two tools for automatically determining replacement policies.
The first tool implements the algorithm proposed in~\cite{Abel13} for inferring permutation policies.
In contrast to the implementation described in~\cite{Abel13}, our implementation is able to determine the policy in individual cache sets. 
The previous implementation assumed that all cache sets use the same policy.

The second tool generates random access sequences, and compares the number of hits obtained by executing them with \emph{cacheSeq} with the number of hits in a simulation of different replacement policies, including common policies like LRU, PLRU, and FIFO, as well as all meaningful QLRU variants, as introduced in~\refsec{bitPol}.

By default, the tool generates random sequences of length~$50$ (which is significantly larger than the associativities of the caches that we consider).
The sequences are generated as follows.
For position $i$ of the sequence, the tool chooses with probability $50\%$ a fresh element and with probability $50\%$ an element that already occurs in the sequence.

We found that typically a relatively small number of such sequences suffices for identifying the correct policy.
In our experiments, evaluating $250$ sequences always produced multiple counterexamples for all but at most one policy.

\subsubsection{Age Graphs}
This tool generates a graph showing the ``ages'' of all blocks of an access sequence that is supplied as a parameter to the tool.
This graph is obtained as follows.
For each block $B$ of the access sequence, we first execute the access sequence, then we access $n$ fresh blocks, and finally we measure the number of hits when accessing $B$ again. %\todo{why is the number of hits not 0 or 1 in~\reffig{ivb}? was the experiment repeated 64 times? - it was performed in all of the cache sets 768-831}
An example of such a graph is discussed in~\refsec{L3Results}.

These graphs are, in particular, useful for analyzing caches with policies that are nondeterministic, and thus cannot be inferred with the tools described above.

\subsubsection{Tests for Set Dueling}
To find the sets with a fixed policy in caches that use set dueling, we implemented an approach similar to~\cite{wong13}.
However, unlike their approach, our tool also supports caches in which the fixed sets are not the same in all C-Boxes.

\begin{table*}
\caption{Replacement Policies Used by Recent Intel CPUs}
%\figurecaptionspace
%\figurecaptionspace
\labeltab{results}
\begin{center}
\begin{tabular}{l|ccc|ccc|ccc}
\toprule
	& \multicolumn{3}{c|}{L1 Data} & \multicolumn{3}{c|}{L2} & \multicolumn{3}{c}{L3}\\
	CPU (Microarchitecture) & Size & \hspace{-1mm}Assoc.\hspace{-1mm} & Policy & Size & \hspace{-1mm}Assoc.\hspace{-1mm} & Policy & Size & \hspace{-1mm}Assoc.\hspace{-1mm} & Policy \\
\midrule
  Core i5-750 (Nehalem)        & 32 kB & 8 & \emph{PLRU}
                               & 256 kB & 8 & \emph{PLRU} 
                               & 8 MB & 16 & \emph{MRU}\\
  Core i5-650 (Westmere)       & 32 kB & 8 & \emph{PLRU} 
                               & 256 kB & 8 & \emph{PLRU} 
                               & 4 MB & 16 & \emph{MRU}\\
  Core i7-2600 (Sandy Bridge)  & 32 kB & 8 & \emph{PLRU} 
                               & 256 kB & 8 & \emph{PLRU} 
                               & 8 MB & 16 & \emph{MRU*}\\
  Core i5-3470 (Ivy Bridge)    & 32 kB & 8 & \emph{PLRU} 
                               & 256 kB & 8 & \emph{PLRU} 
                               & 6 MB & 12  & see~\refsec{L3Results}\\
  Xeon E3-1225 v3 (Haswell)    & 32 kB & 8 & \emph{PLRU} 
                               & 256 kB & 8 & \emph{PLRU} 
                               & 8 MB & 16  & see~\refsec{L3Results}\\
  Core i5-5200U (Broadwell)    & 32 kB & 8 & \emph{PLRU} 
                               & 256 kB & 8 & \emph{PLRU} 
                               & 3 MB & 12  & see~\refsec{L3Results}\\
  Core i7-6500U (Skylake)      & 32 kB & 8 & \emph{PLRU} 
                               & 256 kB & 4 & \emph{QLRU\_H00\_M1\_R2\_U1} 
                               & 4 MB & 16 & see~\refsec{L3Results}\\
  Core i7-7700 (Kaby Lake)     & 32 kB & 8 & \emph{PLRU}
  							   & 256 kB & 4 & \emph{QLRU\_H00\_M1\_R2\_U1} 
  							   & 8 MB & 16 & see~\refsec{L3Results}\\
  Core i7-8700K (Coffee Lake)  & 32 kB & 8 & \emph{PLRU}
                               & 256 kB & 4 & \emph{QLRU\_H00\_M1\_R2\_U1} 
                               & 8 MB & 16 & see~\refsec{L3Results}\\
  Core i3-8121U (Cannon Lake)  & 32 kB & 8 & \emph{PLRU} 
                               & 256 kB & 4 & \emph{QLRU\_H00\_M1\_R0\_U1} 
                               & 4 MB & 16 & see~\refsec{L3Results}\\
  Core i5-1035G1 (Ice Lake)    & 48 kB & 12 & \LRUPLRU
                               & 512 kB & 8  & \emph{QLRU\_H00\_M1\_R0\_U1}
                               & 6 MB & 12 & see~\refsec{L3Results}\\
 \bottomrule
 \end{tabular}
\end{center}
%\figurecaptionspace
\end{table*}

\begin{figure}
\begin{center}
$\begin{array}{rcl}
	\Pi_0 &=& (0, 1, 2, 3, 4, 5, 6, 7, 8, 9, 10, 11)\\
	\Pi_1 &=& (1, 0, 2, 4, 3, 5, 7, 6, 8, 10, 9, 11)\\
	\Pi_2 &=& (2, 0, 1, 5, 3, 4, 8, 6, 7, 11, 9, 10)\\
	\Pi_3 &=& (3, 1, 2, 0, 4, 5, 9, 7, 8, 6, 10, 11)\\
	\Pi_4 &=& (4, 0, 2, 1, 3, 5, 10, 6, 8, 7, 9, 11)\\
	\Pi_5 &=& (5, 0, 1, 2, 3, 4, 11, 6, 7, 8, 9, 10)\\
	\Pi_6 &=& (6, 1, 2, 3, 4, 5, 0, 7, 8, 9, 10, 11)\\
	\Pi_7 &=& (7, 0, 2, 4, 3, 5, 1, 6, 8, 10, 9, 11)\\
	\Pi_8 &=& (8, 0, 1, 5, 3, 4, 2, 6, 7, 11, 9, 10)\\
	\Pi_9 &=& (9, 1, 2, 0, 4, 5, 3, 7, 8, 6, 10, 11)\\
	\Pi_{10} &=& (10, 0, 2, 1, 3, 5, 4, 6, 8, 7, 9, 11)\\
	\Pi_{11} &=& (11, 0, 1, 2, 3, 4, 5, 6, 7, 8, 9, 10)\\
\end{array}$
\end{center}
\caption{Permutation Vectors for the Ice Lake L1 Policy}
\labfig{IceLakeL1}
\end{figure}

\newcommand{\treeNode}[3]{\node[circle, draw, inner sep=.5mm, above=of $(#1)!0.5!(#2)$](#1#2){#3};
  \ifnum#3=0 \draw [->] (#1#2) edge (#1); \draw [-] (#1#2) edge (#2); \fi 
  \ifnum#3=1 \draw [-] (#1#2) edge (#1); \draw [->] (#1#2) edge (#2); \fi}

\begin{figure}
\centering
\begin{tikzpicture}[inner sep=.5mm, node distance=7mm]
\node[draw,minimum size=15pt](n0){$l_0$};
\foreach \n in {1,...,11} {
  \pgfmathparse{int(\n-1)};
  \node[draw,minimum size=15pt,right=0cm of n\pgfmathresult](n\n){$l_{\n}$};
}
\treeNode{n0}{n1}{1};
\treeNode{n2}{n3}{0};
\treeNode{n4}{n5}{1};
\treeNode{n6}{n7}{0};
\treeNode{n8}{n9}{0};
\treeNode{n10}{n11}{1};
\treeNode{n0n1}{n2n3}{0};
\treeNode{n4n5}{n6n7}{1};
\treeNode{n8n9}{n10n11}{0};
\node[ellipse, draw, inner sep=.7mm, above=14mm of n4n5n6n7](topNode){...};
\draw [->, sloped,anchor=south,font=\scriptsize] (topNode) edge node {LRU} (n0n1n2n3);
\draw [->, dotted, sloped,anchor=south,font=\scriptsize] (topNode) edge node {MRU} (n4n5n6n7);
\draw [->, dashed, sloped,anchor=south,font=\scriptsize] (topNode) edge  (n8n9n10n11);
\end{tikzpicture}
\caption{Possible \LRUPLRU State After an Access to $l_4$.}\labfig{IceLakeL1Example}
\end{figure}
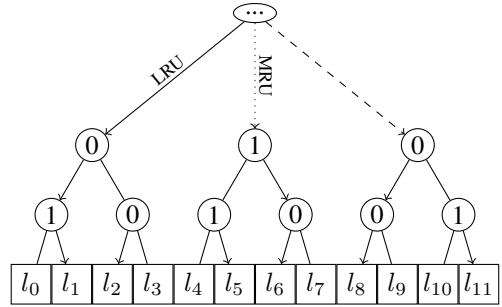

\subsection{Results}\labsec{cacheResults}
We have applied our tools for determining the replacement policies to the Intel CPUs shown in~\reftab{results}.
We did not consider recent AMD CPUs for this case study, as we could not find a way to disable their cache prefetchers, which is required for our cache microbenchmarks (see also \refsec{reducingInterference}).
\medskip

\subsubsection{L1 Data Caches}
The L1 data caches of all processor we considered, except for the Ice Lake CPU, are 8-way set associative and use the PLRU replacement policy.

The L1 data cache of the Ice Lake CPU is 12-way set associative.
It uses a permutation policy; the corresponding permutation vectors are shown in \reffig{IceLakeL1}.
A possible intuitive explanation of this policy is illustrated in \reffig{IceLakeL1Example}.
It uses three PLRU trees with 4 elements each; the trees are ordered by the recency of the last access to one of their elements.
Upon a cache miss, the element that the bits of the least-recently accessed tree point to is replaced.
In the example in \reffig{IceLakeL1Example}, the next element to be replaced would be $l_1$.
We will call this policy \LRUPLRU.
We are unaware of any previous descriptions of this policy.
However, it can be seen as a generalization of the policy used by the 6-way set associative L1 cache of the Intel Atom D525 described in \cite{Abel13}.
\medskip

\subsubsection{L2 Caches}
The L2 caches of CPUs with the Nehalem, Westmere, Sandy Bridge, Ivy Bridge, Haswell, and Broadwell microarchitectures use the PLRU policy.
The more recent generations use two variants of QLRU replacement.
\medskip

\subsubsection{L3 Caches}\labsec{L3Results}

The Nehalem and Westmere CPUs use the MRU replacement policy in their L3 caches.
This was also reported by~\cite{eklov11}.
The Sandy Bridge CPU uses a variant of this policy that sets all bits to one if the cache is not yet full (i.e., after executing the \emph{WBINVD} instruction).

The more recent generations use adaptive policies with different variants of QLRU replacement.

\medskip
\paragraph{Ivy Bridge}

%combination of color list and linestyles*
\pgfplotscreateplotcyclelist{colorlinestyles*}{%
{red,solid},
{blue,dashed},
{black,loosely dotted},
{yellow!70!black,dashdotted},
{brown,dashdotdotted},
{teal,densely dotted},
{orange,solid},
{violet,dashed},
{cyan,loosely dotted},
{green!70!black,dashdotted},
{magenta,dashdotdotted},
{gray,densely dotted},
{olive,solid},
{pink,dashed},
{purple,loosely dotted},
{lime,dashdotted},
{cyan,dashdotdotted},
{green,densely dotted},
{blue,solid},
{yellow,dashed},
{teal,loosely dotted},
{darkgray,dashdotted},
{orange,dashdotdotted},
{magenta,densely dotted}}

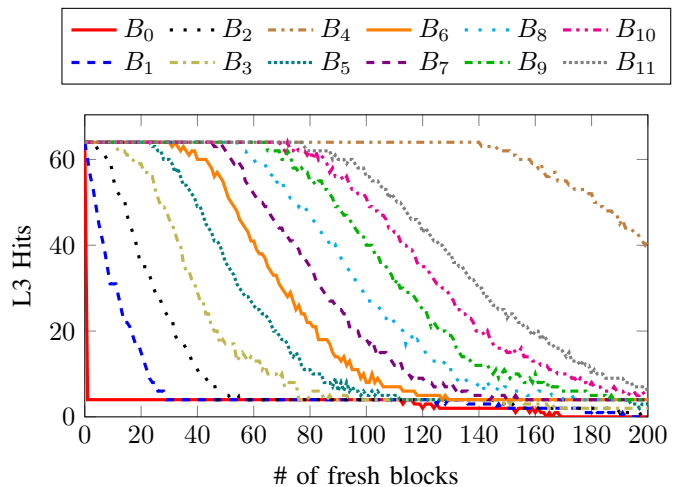
\begin{figure}
\begin{tikzpicture}
\centering
\begin{axis}[width=.5\textwidth, height=5.6cm, cycle list name=colorlinestyles*, xlabel={\# of fresh blocks}, ylabel={L3 Hits}, xmin=0, ymin=0, xmax=200, ylabel near ticks, legend style={at={(0.5,+1.1)},anchor=south, nodes={transform shape}, /tikz/every even column/.append style={column sep=0.1cm}}, transpose legend, legend columns=2, every axis plot/.append style={very thick}]
\addplot table[x=x, y=B0] {ivb_l3.txt}; \addlegendentry{$B_0$}
\addplot table[x=x, y=B1] {ivb_l3.txt}; \addlegendentry{$B_1$}
\addplot table[x=x, y=B2] {ivb_l3.txt}; \addlegendentry{$B_2$}
\addplot table[x=x, y=B3] {ivb_l3.txt}; \addlegendentry{$B_3$}
\addplot table[x=x, y=B4] {ivb_l3.txt}; \addlegendentry{$B_4$}
\addplot table[x=x, y=B5] {ivb_l3.txt}; \addlegendentry{$B_5$}
\addplot table[x=x, y=B6] {ivb_l3.txt}; \addlegendentry{$B_6$}
\addplot table[x=x, y=B7] {ivb_l3.txt}; \addlegendentry{$B_7$}
\addplot table[x=x, y=B8] {ivb_l3.txt}; \addlegendentry{$B_8$}
\addplot table[x=x, y=B9] {ivb_l3.txt}; \addlegendentry{$B_9$}
\addplot table[x=x, y=B10] {ivb_l3.txt}; \addlegendentry{$B_{10}$}
\addplot table[x=x, y=B11] {ivb_l3.txt}; \addlegendentry{$B_{11}$}
\end{axis}
\end{tikzpicture}
\caption{Ivy Bridge Age Graph for the Acc. Seq. ``$\wbinvd\ B_0\,\dots\,B_{11}\ B_{4}$''}
\labfig{ivb}
\end{figure}

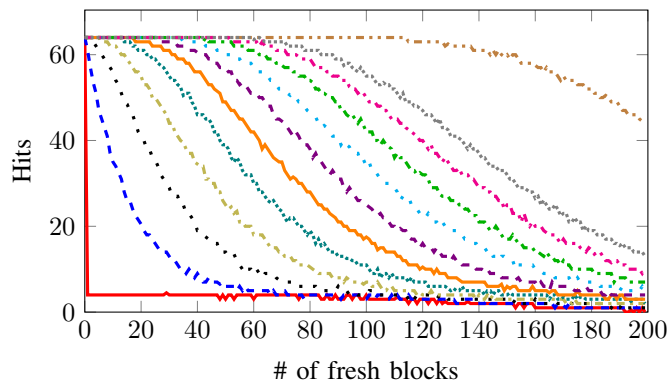
\begin{figure}
\begin{tikzpicture}
\centering
\begin{axis}[width=.5\textwidth, height=5.6cm, cycle list name=colorlinestyles*, xlabel={\# of fresh blocks}, ylabel={Hits}, xmin=0, ymin=0, xmax=200, ylabel near ticks, 
every axis plot/.append style={very thick}]
\addplot table[x=x, y=B0] {ivb_l3_sim.txt}; %\addlegendentry{$B_0$}
\addplot table[x=x, y=B1] {ivb_l3_sim.txt}; %\addlegendentry{$B_1$}
\addplot table[x=x, y=B2] {ivb_l3_sim.txt}; %\addlegendentry{$B_2$}
\addplot table[x=x, y=B3] {ivb_l3_sim.txt}; %\addlegendentry{$B_3$}
\addplot table[x=x, y=B4] {ivb_l3_sim.txt}; %\addlegendentry{$B_4$}
\addplot table[x=x, y=B5] {ivb_l3_sim.txt}; %\addlegendentry{$B_5$}
\addplot table[x=x, y=B6] {ivb_l3_sim.txt}; %\addlegendentry{$B_6$}
\addplot table[x=x, y=B7] {ivb_l3_sim.txt}; %\addlegendentry{$B_7$}
\addplot table[x=x, y=B8] {ivb_l3_sim.txt}; %\addlegendentry{$B_8$}
\addplot table[x=x, y=B9] {ivb_l3_sim.txt}; %\addlegendentry{$B_9$}
\addplot table[x=x, y=B10] {ivb_l3_sim.txt}; %\addlegendentry{$B_{10}$}
\addplot table[x=x, y=B11] {ivb_l3_sim.txt}; %\addlegendentry{$B_{11}$}
\end{axis}
\end{tikzpicture}
\caption{Simulated Age Graph for the Acc. Seq. ``$\wbinvd\ B_0\,\dots\,B_{11}\ B_{4}$''}
\labfig{ivb_sim}
\end{figure}

For the Ivy Bridge machine, we found that the sets 512--575, and the sets 768--831 (in all slices) use a fixed policy, whereas the other sets are follower sets.

According to our results, the sets 512--575 use the \emph{QLRU\_H11\_M1\_R1\_U2} policy.

The policy used by the sets 768--831 appears to be nondeterministic.
\reffig{ivb} shows an age graph for the access sequence ``$\wbinvd\ B_0\ \dots\ B_{11}\ B_{4}$'' (note that the associativity of the cache is $12$); the accesses of the sequence were performed in all of these 64 sets. 
We can see that the curves for $B_i$ and $B_{i+1}$ ($i > 0$) are similar, but shifted by about $16$ (except for~$B4$, which is accessed twice in the sequence).
Furthermore, for $B_0$, about $\frac{15}{16}$ of the blocks are evicted immediately when the first fresh block is accessed, while the remaining $\frac{1}{16}$ of the blocks remain in the cache relatively long.
This suggests that the policy might be the \emph{QLRU\_H11\_MR\textsubscript{16}1\_R1\_U2} policy, i.e., a variant of the policy used in sets 512--575 that inserts new blocks with age $1$ in $\frac{1}{16}th$ of the cases, and with age $3$ otherwise.

\reffig{ivb_sim} shows an age graph for a simulation of this policy. 
The two age graphs are similar, though not identical.
A possible reason for the differences could be that the simulation and the hardware use different random number generators.

\medskip
\paragraph{Haswell, Broadwell}
The Haswell and Broadwell CPUs use the same sets as the Ivy Bridge CPU as dedicated sets, but only in C-Box 0. 
All other sets are follower sets.

Both CPUs use the \emph{QLRU\_H11\_M1\_R0\_U0} policy in sets 512--575.
The policy in sets 768--831 is likely the \emph{QLRU\_H11\_MR\textsubscript{16}1\_R0\_U0} policy.

\medskip
\paragraph{Skylake, Kaby Lake, Coffee Lake, Cannon Lake, Ice Lake}
The Skylake, Kaby Lake, Coffee Lake, and Cannon Lake CPUs use the \emph{QLRU\_H11\_M1\_R0\_U0} policy in 16 fixed cache sets\footnote{Sets 0, 33, 132, 165, 264, 297, 396, 429, 528, 561, 660, 693, 792, 825, 924, and 957}.
The Ice Lake CPU uses the \emph{QLRU\_H00\_M1\_R0\_U1} in the same sets.
These sets were first discovered by Vila et al.~\cite{Vila19a, Vila19b}.

The policy in the remaining sets changes, depending on the number of hits and misses in the sets with the fixed policy, between this policy and a policy that is slightly more thrashing-resistant; unlike with the adaptive policies used by previous generations, there appear to be no sets that always use the second policy.

Further details on all results are available on our website\footnote{\url{www.uops.info/cache.html}}.

\begin{comment}
\section{Experimental Setup}
We ran our tool on the platforms shown in~\reftab{results}.
The machines have between 4 and 32 GB of RAM.
All experiments were performed using Ubuntu~18.04.

For the experiments in Case Study I, we disabled hyperthreading.
For the experiments in Case Study II, we disabled all but one core, and we also disabled cache prefetching.
\todo{in this form, this does not seem to warrant a stand-alone section}
\end{comment}

\section{Related Work}
\subsection{Performance Counter Tools}
\emph{Perf}~\cite{Perf} and Intel's \emph{VTune Amplifier}~\cite{VTune} are two examples of tools that are targeted at analyzing whole programs using hardware performance counters. 
Tools from this category can often display performance statistics at different levels of granularity, sometimes for individual source code lines. 
However, this data is usually obtained via sampling, and thus not precise.
Such tools are commonly used for identifying the parts of a program that would most benefit from further optimizations.

\emph{PAPI}~\cite{PAPI} is a widely used tool for accessing performance counters.
It provides C and Fortran interfaces that provide functions for configuring and reading performance counters.
It can be used for measuring the performance of smaller code segments in the context of a larger program.
However, reading the performance counters leads to multiple function calls, branches, and memory accesses.
Therefore, it is not suitable for the class of microbenchmarks considered in this paper.

\emph{LIKWID}~\cite{Treibig10} is a tool suite providing multiple performance analysis tools.
It can both benchmark whole programs, as well as, similar to \emph{PAPI}, specific code region of a larger program.
Reading the performance counters requires expensive system calls~\cite{likwidOverhead}.

\emph{libpfc}~\cite{libpfc} is a library that was designed in a way to make it possible to use performance counters with a very low overhead.
It provides macros with inline assembler code for reading the performance counters.
Thus, it does not require function calls or branches.
Like our tool, it uses the \emph{LFENCE} instruction to serialize the instruction stream.
In fact, a very early version of our tool was based on \emph{libpfc}.
However, \emph{libpfc} only supports Haswell CPUs, and it does not support accessing uncore performance counters.

Agner Fog~\cite{fogTest} provides a framework for running microbenchmarks similar to the microbenchmarks considered in this paper.
The code of the microbenchmark, which is not allowed to use all registers, must be inserted into specific places in a file provided by the framework.
The overhead for reading performance counters is relatively small; it does not require function calls or branches.
However, the tool uses the \emph{CPUID} instruction for serialization, which can be problematic for short microbenchmarks, as described in \refsec{serialization}.
The tool only supports a relatively small number of performance events, and it only supports performance counters that can be read with the \emph{RDPMC} instruction (i.e., it does not support uncore counters on Intel CPUs, or the APERF/MPERF counters).

In concurrent work, Chen et al.~\cite{Chen19} present a tool for benchmarking basic blocks using the core cycles, and the L1 data and instruction cache performance counters.
Unlike similar tools, Chen et al.'s tool supports microbenchmarks that can make arbitrary memory accesses; this is implemented by automatically mapping all accessed virtual memory pages to a single physical page.
The tool was used to train Ithemal~\cite{mendis19a}, which is a basic block throughput predictor that is based on a neural network.
Chen et al. also propose a benchmark suite, called \emph{BHive}, that consists of more than $300,000$ basic blocks, and they use their tool to obtain throughput measurements for these basic blocks on CPUs with the Ivy Bridge, Haswell, and Skylake microarchitectures.
The code that reads the performance counters contains branches, and it uses the \emph{CPUID} instruction for serialization; however, it lacks a serialization instruction after reading the core cycles counter for the first time.
As a consequence, the measurement results are relatively noisy.
For Skylake, for example, we found in the \emph{BHive} benchmark suite about $20,000$ basic blocks that have instructions with memory operands, but a measured throughput value smaller than $0.5$; for more than $2,200$ of these blocks, the measured throughput value was even smaller than $0.45$\footnote{https://github.com/ithemal/bhive/issues/1}. These throughput values are obviously incorrect, since Skylake can execute at most two instructions with memory operands per cycle.
Chen et al. compare their measurement results with predictions from Ithemal and several other throughput prediction tools. 
As the average deviation of Ithemal's predictions from the measured throughputs is smaller than the average deviations of the other tools, the authors conclude that Ithemal outperforms the other tools.

None of the existing tools that we are aware of allows for executing benchmarks directly in kernel space.

\subsection{Microbenchmark-Based Cache Analysis}
A number of papers have proposed microbenchmark-based techniques for determining parameters of the memory hierarchy like the cache size, the associativity, the block size, or the latency \cite{Saavedra95,Thomborson00,Coleman01,Dongarra04,Yotov05a, Yotov06, Babka09, Molka09, Wong10, Abel12a, Hassan15, Mei17, Cooper18}. 

While some of these approaches make assumptions as to the underlying replacement policy (e.g. \cite{Saavedra95} and~\cite{Thomborson00} assume that LRU is used), only a few publications have also tried to determine the replacement policy. 

The approaches described in~\cite{Coleman01} and~\cite{Blanquer00} are able to detect LRU-based policies but treat all other policies as random. 
John and Baumgartl's~\cite{John07} approach is able to distinguish between LRU and several of its derivatives.

In \cite{Abel13}, we proposed an algorithm that can automatically infer \emph{permutation policies}. 
In the present work, we developed an improved implementation of this algorithm that can infer the policies in individual cache sets; the implementation in~\cite{Abel13} was based on the assumption that all cache sets use the same policy.

Henry Wong~\cite{wong13} discovered that Ivy Bridge CPUs use set dueling to switch between two different replacement policies. 
He identified the sets that use fixed policies; however, he was not able to determine which two policies are actually used.
Similar work was described by Zhang et al.~\cite{Zhang14}.

Briongos et al.~\cite{briongos19} present an approach for analyzing the replacement policies used by the L3 caches in recent Intel CPUs.
Similar to the technique described in \refsec{toolReplPol}, their approach generates random access sequences and compares their behavior to simulations of different policies.
However, unlike in our approach, they do not measure the total number of hits that the sequence generates.
Instead, they only determine the first element to be evicted upon the first miss after executing the access sequence.
Furthermore, unlike in our approach, they rely on timing measurements instead of using performance counters.

Briongos et al. applied their technique to CPUs with the Haswell, Broadwell, Skylake, and Kaby Lake microarchitectures.
Our results for these microarchitectures disagree with their results.
The policies they describe would be the \emph{QLRU\_H21\_M2\_R0\_U0\_UMO} and \emph{QLRU\_H21\_M3\_R0\_U0\_UMO}  variants according to our naming scheme.
Our tool found several counterexamples for these policies on all of the tested CPUs.
Briongos et al. also stated that the two policies did not agree with all of their observations; however, they assumed that ``the errors were due to noise''.
Furthermore, according to Briongos et al., the dedicated sets on the Haswell and Broadwell CPUs are distributed over different slices; according to our results, they are all in the same slice.
As, according to the paper, they use an approach from~\cite{liu15}, we assume that they also rely on a statement from that paper that ``when the number of cores in the processor is a power of two, the set index bits are not used for determining the LLC slice.''
This was, however, shown to be incorrect in later work~\cite{Maurice15}.
Thus, their observations rather seem to be an artifact of the hash function used for determining the cache slices.
Briongos et al. did not find the sets with a varying policy on Skylake and Kaby Lake, and thus incorrectly concluded that these CPUs do not use an adaptive policy.

\begin{comment}
Rueda \cite{Rueda13} developed a technique for learning replacement policies using register automata.
He was able to learn the FIFO and LRU policies for caches with an associativity of at most 5, and the PLRU and MRU policies for caches with an associativity of at most 4.
He did not successfully apply his technique to actual hardware.
\end{comment}

In concurrent work, Vila et al.~\cite{Vila19b} describe an approach for inferring replacement policies using automata learning, and an approach for automatically generating human-readable representations of the learned policies.
For software-simulated caches, they were able to learn FIFO and PLRU up to associativity 16, MRU up to associativity 12, and several other policies up to associativity 6.
Furthermore, Vila et al. also applied their techniques to actual hardware with the Haswell, Skylake, and Kaby Lake microarchitectures.
They successfully learned the policies used by the L1 and L2 caches of these three processors, as well as the policy used by the leader sets of the L3 caches on Skylake and Kaby Lake.
Their results agree with our results.
Vila et al.'s approach was, however, not able to learn the policies used by the L3 cache of the Haswell CPU, as the associativity was too high for their approach, and one of the policies is nondeterministic.

Vila et al.'s approach relies on a tool called \emph{CacheQuery}, that is quite similar to the \emph{CacheSeq} tool proposed in \refsec{cacheTools}. 
The main differences are the following:
\begin{enumerate}
\item \emph{CacheQuery} uses a more expressive syntax,
\item \emph{CacheQuery} is based on timing measurements, whereas \emph{CacheSeq} uses performance counters, and
\item \emph{CacheQuery} requires the parameters of the caches, such as the associativities or the number of cache sets, to be specified manually, whereas \emph{CacheSeq} determines them automatically.
\end{enumerate}

\section{Conclusions and Future Work}

We have presented a new tool that significantly reduces the engineering effort required for evaluating small microbenchmarks in an accurate and precise way.

To illustrate the usefulness of our tool, we have presented two different case studies.
First, we showed how it can be used to characterize the latency, throughput, and port usage
of x86 instructions.
Then, we described microbenchmarks for analyzing cache properties.
We applied these microbenchmarks to recent Intel CPUs, and uncovered several previously undocumented replacement policy variants.

There are two main directions for future work.
The first direction is to adapt \nanoBench to non-x86 architectures, such as ARM.
The second direction is to apply \nanoBench to additional use cases. 
Besides the two examples we considered, many other properties of recent microarchitectures are undocumented.
This includes, for example, details on how the TLBs or the branch predictors work.
Knowledge of such details is important for optimizing software, and for showing the presence or absence of microarchitectural security problems.

\bibliographystyle{IEEEtran}
\bibliography{references}

\begin{thebibliography}{10}
\providecommand{\url}[1]{#1}
\csname url@samestyle\endcsname
\providecommand{\newblock}{\relax}
\providecommand{\bibinfo}[2]{#2}
\providecommand{\BIBentrySTDinterwordspacing}{\spaceskip=0pt\relax}
\providecommand{\BIBentryALTinterwordstretchfactor}{4}
\providecommand{\BIBentryALTinterwordspacing}{\spaceskip=\fontdimen2\font plus
\BIBentryALTinterwordstretchfactor\fontdimen3\font minus
  \fontdimen4\font\relax}
\providecommand{\BIBforeignlanguage}[2]{{%
\expandafter\ifx\csname l@#1\endcsname\relax
\typeout{** WARNING: IEEEtran.bst: No hyphenation pattern has been}%
\typeout{** loaded for the language `#1'. Using the pattern for}%
\typeout{** the default language instead.}%
\else
\language=\csname l@#1\endcsname
\fi
#2}}
\providecommand{\BIBdecl}{\relax}
\BIBdecl

\bibitem{Abel19}
\BIBentryALTinterwordspacing
A.~Abel and J.~Reineke, ``uops.info: Characterizing latency, throughput, and
  port usage of instructions on {Intel} microarchitectures,'' in
  \emph{Proceedings of the Twenty-Fourth International Conference on
  Architectural Support for Programming Languages and Operating Systems}, ser.
  ASPLOS '19.\hskip 1em plus 0.5em minus 0.4em\relax New York, NY, USA: ACM,
  2019, pp. 673--686. [Online]. Available:
  \url{http://doi.acm.org/10.1145/3297858.3304062}
\BIBentrySTDinterwordspacing

\bibitem{cqa14}
\BIBentryALTinterwordspacing
A.~S. Charif-Rubial, E.~Oseret, J.~Noudohouenou, W.~Jalby, and G.~Lartigue,
  ``{CQA}: A code quality analyzer tool at binary level,'' in \emph{21st
  International Conference on High Performance Computing (HiPC)}, Dec 2014, pp.
  1--10. [Online]. Available:
  \url{http://www.maqao.org/publications/papers/CQA.pdf}
\BIBentrySTDinterwordspacing

\bibitem{fog17}
\BIBentryALTinterwordspacing
A.~Fog, \emph{Instruction tables: Lists of instruction latencies, throughputs
  and micro-operation breakdowns for {Intel}, {AMD} and {VIA CPUs}}, Technical
  University of Denmark, May 2017. [Online]. Available:
  \url{http://www.agner.org/optimize/instruction_tables.pdf}
\BIBentrySTDinterwordspacing

\bibitem{granlund17}
\BIBentryALTinterwordspacing
T.~Granlund, ``Instruction latencies and throughput for {AMD} and {Intel} x86
  processors,'' Apr. 2017. [Online]. Available:
  \url{https://gmplib.org/~tege/x86-timing.pdf}
\BIBentrySTDinterwordspacing

\bibitem{instlatx64}
\BIBentryALTinterwordspacing
 [Online]. Available: \url{http://instlatx64.atw.hu/}
\BIBentrySTDinterwordspacing

\bibitem{Saavedra95}
\BIBentryALTinterwordspacing
R.~H. Saavedra and A.~J. Smith, ``Measuring cache and {TLB} performance and
  their effect on benchmark runtimes,'' \emph{{IEEE} Trans. Computers},
  vol.~44, no.~10, pp. 1223--1235, 1995. [Online]. Available:
  \url{https://doi.org/10.1109/12.467697}
\BIBentrySTDinterwordspacing

\bibitem{Thomborson00}
\BIBentryALTinterwordspacing
C.~Thomborson and Y.~Yu, ``Measuring data cache and {TLB} parameters under
  {Linux},'' in \emph{Proceedings of the Symposium on Performance Evaluation of
  Computer and Telecommunication Systems}, Jul. 2000, pp. 383--390. [Online].
  Available:
  \url{http://citeseerx.ist.psu.edu/viewdoc/summary?doi=10.1.1.36.1427}
\BIBentrySTDinterwordspacing

\bibitem{Coleman01}
\BIBentryALTinterwordspacing
C.~Coleman and J.~Davidson, ``Automatic memory hierarchy characterization,'' in
  \emph{ISPASS}, 2001, pp. 103--110. [Online]. Available:
  \url{http://dx.doi.org/10.1109/ISPASS.2001.990684}
\BIBentrySTDinterwordspacing

\bibitem{Dongarra04}
\BIBentryALTinterwordspacing
J.~J. Dongarra, S.~Moore, P.~Mucci, K.~Seymour, and H.~You, ``Accurate cache
  and {TLB} characterization using hardware counters,'' in \emph{ICCS}, 2004,
  pp. 432--439. [Online]. Available:
  \url{http://dx.doi.org/10.1007/978-3-540-24688-6_57}
\BIBentrySTDinterwordspacing

\bibitem{Yotov05a}
\BIBentryALTinterwordspacing
K.~Yotov, K.~Pingali, and P.~Stodghill, ``Automatic measurement of memory
  hierarchy parameters,'' in \emph{SIGMETRICS}.\hskip 1em plus 0.5em minus
  0.4em\relax New York, NY, USA: ACM, 2005, pp. 181--192. [Online]. Available:
  \url{http://doi.acm.org/10.1145/1064212.1064233}
\BIBentrySTDinterwordspacing

\bibitem{Yotov06}
\BIBentryALTinterwordspacing
K.~Yotov, S.~Jackson, T.~Steele, K.~Pingali, and P.~Stodghill, ``Automatic
  measurement of instruction cache capacity,'' in \emph{Proceedings of the 18th
  international workshop on Languages and Compilers for Parallel
  Computing}.\hskip 1em plus 0.5em minus 0.4em\relax Springer, 2006, pp.
  230--243. [Online]. Available:
  \url{http://dx.doi.org/10.1007/978-3-540-69330-7_16}
\BIBentrySTDinterwordspacing

\bibitem{Molka09}
\BIBentryALTinterwordspacing
D.~Molka, D.~Hackenberg, R.~Sch{\"{o}}ne, and M.~S. M{\"{u}}ller, ``Memory
  performance and cache coherency effects on an {Intel} {Nehalem}
  multiprocessor system,'' in \emph{Proceedings of the 2009 18th International
  Conference on Parallel Architectures and Compilation Techniques}, ser. PACT
  '09.\hskip 1em plus 0.5em minus 0.4em\relax Washington, DC, USA: IEEE, 2009,
  pp. 261--270. [Online]. Available:
  \url{http://dx.doi.org/10.1109/PACT.2009.22}
\BIBentrySTDinterwordspacing

\bibitem{Babka09}
\BIBentryALTinterwordspacing
V.~Babka and P.~T\r{u}ma, ``Investigating cache parameters of x86 family
  processors,'' in \emph{Proceedings of the 2009 SPEC benchmark
  workshop}.\hskip 1em plus 0.5em minus 0.4em\relax Springer, 2009, pp. 77--96.
  [Online]. Available: \url{http://dx.doi.org/10.1007/978-3-540-93799-9_5}
\BIBentrySTDinterwordspacing

\bibitem{Wong10}
\BIBentryALTinterwordspacing
H.~Wong, M.-M. Papadopoulou, M.~Sadooghi-Alvandi, and A.~Moshovos,
  ``Demystifying {GPU} microarchitecture through microbenchmarking,'' in
  \emph{ISPASS}, 2010, pp. 235--246. [Online]. Available:
  \url{http://dx.doi.org/10.1109/ISPASS.2010.5452013}
\BIBentrySTDinterwordspacing

\bibitem{Abel12a}
\BIBentryALTinterwordspacing
A.~Abel and J.~Reineke, ``Automatic cache modeling by measurements,'' in
  \emph{6th Junior Researcher Workshop on Real-Time Computing (in conjunction
  with RTNS)}, Nov. 2012. [Online]. Available:
  \url{http://embedded.cs.uni-saarland.de/publications/CacheModelingJRWRTC.pdf}
\BIBentrySTDinterwordspacing

\bibitem{Abel13}
\BIBentryALTinterwordspacing
------, ``Measurement-based modeling of the cache replacement policy,'' in
  \emph{19th {IEEE} Real-Time and Embedded Technology and Applications
  Symposium, {RTAS}, Philadelphia, PA, USA}, 2013, pp. 65--74. [Online].
  Available: \url{https://doi.org/10.1109/RTAS.2013.6531080}
\BIBentrySTDinterwordspacing

\bibitem{Abel14}
\BIBentryALTinterwordspacing
------, ``Reverse engineering of cache replacement policies in {Intel}
  microprocessors and their evaluation,'' in \emph{2014 {IEEE} International
  Symposium on Performance Analysis of Systems and Software, {ISPASS} 2014,
  Monterey, CA, USA, March 23-25, 2014}, 2014, pp. 141--142. [Online].
  Available: \url{https://doi.org/10.1109/ISPASS.2014.6844475}
\BIBentrySTDinterwordspacing

\bibitem{Hassan15}
\BIBentryALTinterwordspacing
M.~Hassan, A.~M. Kaushik, and H.~D. Patel, ``Reverse-engineering embedded
  memory controllers through latency-based analysis,'' in \emph{21st {IEEE}
  Real-Time and Embedded Technology and Applications Symposium, Seattle, WA,
  USA}, 2015, pp. 297--306. [Online]. Available:
  \url{https://doi.org/10.1109/RTAS.2015.7108453}
\BIBentrySTDinterwordspacing

\bibitem{Mei17}
\BIBentryALTinterwordspacing
X.~Mei and X.~Chu, ``Dissecting {GPU} memory hierarchy through
  microbenchmarking,'' \emph{{IEEE} Trans. Parallel Distrib. Syst.}, vol.~28,
  no.~1, pp. 72--86, 2017. [Online]. Available:
  \url{https://doi.org/10.1109/TPDS.2016.2549523}
\BIBentrySTDinterwordspacing

\bibitem{Cooper18}
\BIBentryALTinterwordspacing
K.~Cooper and X.~Xu, ``Efficient characterization of hidden processor memory
  hierarchies,'' in \emph{Computational Science - {ICCS} 2018 - 18th
  International Conference, Wuxi, China, June 11-13, 2018 Proceedings, Part
  {III}}, 2018, pp. 335--349. [Online]. Available:
  \url{http://dx.doi.org/10.1007/978-3-319-93713-7_27}
\BIBentrySTDinterwordspacing

\bibitem{kocher19}
\BIBentryALTinterwordspacing
P.~Kocher, J.~Horn, A.~Fogh, D.~Genkin, D.~Gruss, W.~Haas, M.~Hamburg, M.~Lipp,
  S.~Mangard, T.~Prescher, and et~al., ``Spectre attacks: Exploiting
  speculative execution,'' \emph{2019 IEEE Symposium on Security and Privacy
  (SP)}, May 2019. [Online]. Available:
  \url{http://dx.doi.org/10.1109/sp.2019.00002}
\BIBentrySTDinterwordspacing

\bibitem{lipp18}
\BIBentryALTinterwordspacing
M.~Lipp, M.~Schwarz, D.~Gruss, T.~Prescher, W.~Haas, S.~Mangard, P.~Kocher,
  D.~Genkin, Y.~Yarom, and M.~Hamburg, ``Meltdown,'' 2018. [Online]. Available:
  \url{https://arxiv.org/pdf/1801.01207.pdf}
\BIBentrySTDinterwordspacing

\bibitem{Perf}
\BIBentryALTinterwordspacing
``perf: Linux profiling with performance counters.'' [Online]. Available:
  \url{https://perf.wiki.kernel.org}
\BIBentrySTDinterwordspacing

\bibitem{VTune}
\BIBentryALTinterwordspacing
``{Intel VTune Amplifier}.'' [Online]. Available:
  \url{https://software.intel.com/vtune}
\BIBentrySTDinterwordspacing

\bibitem{PAPI}
\BIBentryALTinterwordspacing
D.~Terpstra, H.~Jagode, H.~You, and J.~Dongarra, ``Collecting performance data
  with {PAPI-C},'' in \emph{Tools for High Performance Computing 2009}.\hskip
  1em plus 0.5em minus 0.4em\relax Springer, 2010, pp. 157--173. [Online].
  Available: \url{http://dx.doi.org/10.1007/978-3-642-11261-4_11}
\BIBentrySTDinterwordspacing

\bibitem{libpfc}
\BIBentryALTinterwordspacing
``libpfc.'' [Online]. Available: \url{https://github.com/obilaniu/libpfc}
\BIBentrySTDinterwordspacing

\bibitem{intelOptManual19}
\BIBentryALTinterwordspacing
\emph{Intel 64 and {IA-32} Architectures Optimization Reference Manual}, Intel
  Corporation, Sep. 2019, {O}rder Number: 248966-042b. [Online]. Available:
  \url{https://software.intel.com/sites/default/files/managed/9e/bc/64-ia-32-architectures-optimization-manual.pdf}
\BIBentrySTDinterwordspacing

\bibitem{fogTest}
\BIBentryALTinterwordspacing
A.~Fog, ``Test programs for measuring clock cycles and performance
  monitoring.'' [Online]. Available: \url{https://www.agner.org/optimize/}
\BIBentrySTDinterwordspacing

\bibitem{paoloni2010benchmark}
\BIBentryALTinterwordspacing
G.~Paoloni, ``How to benchmark code execution times on {Intel IA-32} and
  {IA-64} instruction set architectures,'' \emph{Intel Corporation}, September
  2010. [Online]. Available:
  \url{https://www.intel.com/content/dam/www/public/us/en/documents/white-papers/ia-32-ia-64-benchmark-code-execution-paper.pdf}
\BIBentrySTDinterwordspacing

\bibitem{intelDevManual}
\BIBentryALTinterwordspacing
\emph{Intel{\textregistered} 64 and {IA-32} Architectures Software Developer's
  Manual, Volume 2}, Intel Corporation, Nov. 2018, {O}rder Number:
  325383-068US. [Online]. Available:
  \url{https://software.intel.com/sites/default/files/managed/a4/60/325383-sdm-vol-2abcd.pdf}
\BIBentrySTDinterwordspacing

\bibitem{McCalpin18}
\BIBentryALTinterwordspacing
J.~D. McCalpin, ``Comments on timing short code sections on {Intel}
  processors.'' [Online]. Available:
  \url{https://sites.utexas.edu/jdm4372/2018/07/23/comments-on-timing-short-code-sections-on-intel-processors/}
\BIBentrySTDinterwordspacing

\bibitem{weaver2008can}
\BIBentryALTinterwordspacing
V.~M. Weaver and S.~A. McKee, ``Can hardware performance counters be trusted?''
  in \emph{2008 IEEE International Symposium on Workload
  Characterization}.\hskip 1em plus 0.5em minus 0.4em\relax IEEE, Sep. 2008,
  pp. 141--150. [Online]. Available:
  \url{http://dx.doi.org/10.1109/IISWC.2008.4636099}
\BIBentrySTDinterwordspacing

\bibitem{weaver2013non}
\BIBentryALTinterwordspacing
V.~M. Weaver, D.~Terpstra, and S.~Moore, ``Non-determinism and overcount on
  modern hardware performance counter implementations,'' in \emph{2013 IEEE
  International Symposium on Performance Analysis of Systems and Software
  (ISPASS)}.\hskip 1em plus 0.5em minus 0.4em\relax IEEE, Apr. 2013, pp.
  215--224. [Online]. Available:
  \url{http://dx.doi.org/10.1109/ISPASS.2013.6557172}
\BIBentrySTDinterwordspacing

\bibitem{hund13}
\BIBentryALTinterwordspacing
R.~Hund, C.~Willems, and T.~Holz, ``Practical timing side channel attacks
  against kernel space {ASLR},'' in \emph{2013 IEEE Symposium on Security and
  Privacy}.\hskip 1em plus 0.5em minus 0.4em\relax IEEE, 2013, pp. 191--205.
  [Online]. Available: \url{http://dx.doi.org/10.1109/SP.2013.23}
\BIBentrySTDinterwordspacing

\bibitem{irazoqui15}
\BIBentryALTinterwordspacing
G.~Irazoqui, T.~Eisenbarth, and B.~Sunar, ``Systematic reverse engineering of
  cache slice selection in {Intel} processors,'' in \emph{2015 Euromicro
  Conference on Digital System Design}.\hskip 1em plus 0.5em minus 0.4em\relax
  IEEE, 2015, pp. 629--636. [Online]. Available:
  \url{http://dx.doi.org/10.1109/DSD.2015.56}
\BIBentrySTDinterwordspacing

\bibitem{liu15}
\BIBentryALTinterwordspacing
F.~Liu, Y.~Yarom, Q.~Ge, G.~Heiser, and R.~B. Lee, ``Last-level cache
  side-channel attacks are practical,'' in \emph{Proceedings of the 2015 IEEE
  Symposium on Security and Privacy}.\hskip 1em plus 0.5em minus 0.4em\relax
  IEEE, 2015, pp. 605--622. [Online]. Available:
  \url{http://dx.doi.org/10.1109/SP.2015.43}
\BIBentrySTDinterwordspacing

\bibitem{Maurice15}
\BIBentryALTinterwordspacing
C.~Maurice, N.~Scouarnec, C.~Neumann, O.~Heen, and A.~Francillon, ``Reverse
  engineering {Intel} last-level cache complex addressing using performance
  counters,'' in \emph{Proceedings of the 18th International Symposium on
  Research in Attacks, Intrusions, and Defenses - Volume 9404}, ser. RAID
  2015.\hskip 1em plus 0.5em minus 0.4em\relax New York, NY, USA:
  Springer-Verlag, 2015, pp. 48--65. [Online]. Available:
  \url{http://dx.doi.org/10.1007/978-3-319-26362-5_3}
\BIBentrySTDinterwordspacing

\bibitem{yarom15}
\BIBentryALTinterwordspacing
Y.~Yarom, Q.~Ge, F.~Liu, R.~B. Lee, and G.~Heiser, ``Mapping the {Intel}
  last-level cache.'' \emph{Cryptology ePrint Archive, Report 2015/905}, 2015.
  [Online]. Available: \url{https://eprint.iacr.org/2015/905}
\BIBentrySTDinterwordspacing

\bibitem{inci16}
\BIBentryALTinterwordspacing
M.~S. Inci, B.~Gulmezoglu, G.~Irazoqui, T.~Eisenbarth, and B.~Sunar, ``Cache
  attacks enable bulk key recovery on the cloud,'' in \emph{International
  Conference on Cryptographic Hardware and Embedded Systems}.\hskip 1em plus
  0.5em minus 0.4em\relax Springer, 2016, pp. 368--388. [Online]. Available:
  \url{http://dx.doi.org/10.1007/978-3-662-53140-2_18}
\BIBentrySTDinterwordspacing

\bibitem{kayaalp16}
\BIBentryALTinterwordspacing
M.~Kayaalp, N.~Abu-Ghazaleh, D.~Ponomarev, and A.~Jaleel, ``A high-resolution
  side-channel attack on last-level cache,'' in \emph{Proceedings of the 53rd
  Annual Design Automation Conference}.\hskip 1em plus 0.5em minus 0.4em\relax
  ACM, 2016. [Online]. Available:
  \url{http://dx.doi.org/10.1145/2897937.2897962}
\BIBentrySTDinterwordspacing

\bibitem{Disselkoen17}
\BIBentryALTinterwordspacing
C.~Disselkoen, D.~Kohlbrenner, L.~Porter, and D.~Tullsen, ``Prime+abort: A
  timer-free high-precision {L3} cache attack using {Intel TSX},'' in
  \emph{Proceedings of the 26th USENIX Conference on Security Symposium}, ser.
  SEC'17.\hskip 1em plus 0.5em minus 0.4em\relax Berkeley, CA, USA: USENIX
  Association, 2017, pp. 51--67. [Online]. Available:
  \url{http://dl.acm.org/citation.cfm?id=3241189.3241195}
\BIBentrySTDinterwordspacing

\bibitem{Farshin19}
\BIBentryALTinterwordspacing
A.~Farshin, A.~Roozbeh, G.~Q. Maguire, Jr., and D.~Kosti\'{c}, ``Make the most
  out of last level cache in {Intel} processors,'' in \emph{Proceedings of the
  Fourteenth EuroSys Conference 2019}, ser. EuroSys '19.\hskip 1em plus 0.5em
  minus 0.4em\relax New York, NY, USA: ACM, 2019, pp. 8:1--8:17. [Online].
  Available: \url{http://doi.acm.org/10.1145/3302424.3303977}
\BIBentrySTDinterwordspacing

\bibitem{reineke07}
\BIBentryALTinterwordspacing
J.~Reineke, D.~Grund, C.~Berg, and R.~Wilhelm, ``Timing predictability of cache
  replacement policies,'' \emph{Real-Time Systems}, vol.~37, no.~2, pp.
  99--122, Nov 2007. [Online]. Available:
  \url{https://doi.org/10.1007/s11241-007-9032-3}
\BIBentrySTDinterwordspacing

\bibitem{pan15}
\BIBentryALTinterwordspacing
X.~Pan and B.~Jonsson, ``A modeling framework for reuse distance-based
  estimation of cache performance,'' in \emph{2015 IEEE International Symposium
  on Performance Analysis of Systems and Software (ISPASS)}.\hskip 1em plus
  0.5em minus 0.4em\relax IEEE, Mar. 2015, pp. 62--71. [Online]. Available:
  \url{http://dx.doi.org/10.1109/ISPASS.2015.7095785}
\BIBentrySTDinterwordspacing

\bibitem{al04}
\BIBentryALTinterwordspacing
H.~Al-Zoubi, A.~Milenkovic, and M.~Milenkovic, ``Performance evaluation of
  cache replacement policies for the {SPEC} {CPU2000} benchmark suite,'' in
  \emph{Proceedings of the 42nd annual Southeast regional conference}.\hskip
  1em plus 0.5em minus 0.4em\relax ACM, 2004, pp. 267--272. [Online].
  Available: \url{http://dx.doi.org/10.1145/986537.986601}
\BIBentrySTDinterwordspacing

\bibitem{jaleel10}
\BIBentryALTinterwordspacing
A.~Jaleel, K.~B. Theobald, S.~C. Steely, Jr., and J.~Emer, ``High performance
  cache replacement using re-reference interval prediction ({RRIP}),'' in
  \emph{Proceedings of the 37th Annual International Symposium on Computer
  Architecture}, ser. ISCA '10.\hskip 1em plus 0.5em minus 0.4em\relax New
  York, NY, USA: ACM, 2010, pp. 60--71. [Online]. Available:
  \url{http://doi.acm.org/10.1145/1815961.1815971}
\BIBentrySTDinterwordspacing

\bibitem{jahagirdar12}
\BIBentryALTinterwordspacing
S.~Jahagirdar, V.~George, I.~Sodhi, and R.~Wells, ``Power management of the
  third generation {Intel Core} micro architecture formerly codenamed {Ivy
  Bridge},'' \emph{2012 IEEE Hot Chips 24 Symposium (HCS)}, Aug 2012. [Online].
  Available: \url{http://dx.doi.org/10.1109/hotchips.2012.7476478}
\BIBentrySTDinterwordspacing

\bibitem{briongos19}
\BIBentryALTinterwordspacing
S.~Briongos, P.~Malag\'{o}n, J.~M. Moya, and T.~Eisenbarth, ``Reload+refresh:
  Abusing cache replacement policies to perform stealthy cache attacks,'' 2019.
  [Online]. Available: \url{https://arxiv.org/abs/1904.06278}
\BIBentrySTDinterwordspacing

\bibitem{qureshi07}
\BIBentryALTinterwordspacing
M.~K. Qureshi, A.~Jaleel, Y.~N. Patt, S.~C. Steely, and J.~Emer, ``Adaptive
  insertion policies for high performance caching,'' in \emph{Proceedings of
  the 34th Annual International Symposium on Computer Architecture}, ser. ISCA
  '07.\hskip 1em plus 0.5em minus 0.4em\relax New York, NY, USA: ACM, 2007, pp.
  381--391. [Online]. Available:
  \url{http://doi.acm.org/10.1145/1250662.1250709}
\BIBentrySTDinterwordspacing

\bibitem{wong13}
\BIBentryALTinterwordspacing
H.~Wong, ``{Intel Ivy Bridge} cache replacement policy,'' 2013. [Online].
  Available: \url{http://blog.stuffedcow.net/2013/01/ivb-cache-replacement/}
\BIBentrySTDinterwordspacing

\bibitem{Vila19b}
\BIBentryALTinterwordspacing
P.~Vila, P.~Ganty, M.~Guarnieri, and B.~K{\"{o}}pf, ``{CacheQuery}: Learning
  replacement policies from hardware caches,'' \emph{CoRR}, vol.
  abs/1912.09770, Dec. 2019. [Online]. Available:
  \url{http://arxiv.org/abs/1912.09770}
\BIBentrySTDinterwordspacing

\bibitem{eklov11}
\BIBentryALTinterwordspacing
D.~Eklov, N.~Nikoleris, D.~Black-Schaffer, and E.~Hagersten, ``Cache pirating:
  Measuring the curse of the shared cache,'' in \emph{Proceedings of the 2011
  International Conference on Parallel Processing}, ser. ICPP '11.\hskip 1em
  plus 0.5em minus 0.4em\relax Washington, DC, USA: IEEE Computer Society,
  2011, pp. 165--175. [Online]. Available:
  \url{http://dx.doi.org/10.1109/ICPP.2011.15}
\BIBentrySTDinterwordspacing

\bibitem{Vila19a}
\BIBentryALTinterwordspacing
P.~Vila, B.~K{\"{o}}pf, and J.~F. Morales, ``Theory and practice of finding
  eviction sets,'' in \emph{2019 {IEEE} Symposium on Security and Privacy
  ({SP})}, May 2019, pp. 39--54. [Online]. Available:
  \url{http://dx.doi.org/10.1109/SP.2019.00042}
\BIBentrySTDinterwordspacing

\bibitem{Treibig10}
\BIBentryALTinterwordspacing
J.~Treibig, G.~Hager, and G.~Wellein, ``{LIKWID}: A lightweight
  performance-oriented tool suite for x86 multicore environments,'' in
  \emph{Proceedings of the International Conference on Parallel Processing
  Workshops}.\hskip 1em plus 0.5em minus 0.4em\relax Washington, DC, USA: IEEE,
  2010, pp. 207--216. [Online]. Available:
  \url{http://dx.doi.org/10.1109/ICPPW.2010.38}
\BIBentrySTDinterwordspacing

\bibitem{likwidOverhead}
\BIBentryALTinterwordspacing
T.~Roehl, J.~Treibig, G.~Hager, and G.~Wellein, ``Overhead analysis of
  performance counter measurements,'' in \emph{43rd International Conference on
  Parallel Processing Workshops (ICCPW)}, Sep. 2014, pp. 176--185. [Online].
  Available: \url{http://dx.doi.org/10.1109/ICPPW.2014.34}
\BIBentrySTDinterwordspacing

\bibitem{Chen19}
\BIBentryALTinterwordspacing
Y.~Chen, A.~Brahmakshatriya, C.~Mendis, A.~Renda, E.~Atkinson, O.~Sykora,
  S.~Amarasinghe, and M.~Carbin, ``{BHive}: A benchmark suite and measurement
  framework for validating x86-64 basic block performance models,'' in
  \emph{2019 {IEEE} International Symposium on Workload Characterization
  (IISWC)}.\hskip 1em plus 0.5em minus 0.4em\relax IEEE, Nov. 2019. [Online].
  Available:
  \url{http://groups.csail.mit.edu/commit/papers/19/ithemal-measurement.pdf}
\BIBentrySTDinterwordspacing

\bibitem{mendis19a}
\BIBentryALTinterwordspacing
C.~Mendis, A.~Renda, D.~Amarasinghe, and M.~Carbin, ``Ithemal: Accurate,
  portable and fast basic block throughput estimation using deep neural
  networks,'' in \emph{Proceedings of the 36th International Conference on
  Machine Learning}, ser. Proceedings of Machine Learning Research,
  K.~Chaudhuri and R.~Salakhutdinov, Eds., vol.~97.\hskip 1em plus 0.5em minus
  0.4em\relax Long Beach, California, USA: PMLR, 09--15 Jun 2019, pp.
  4505--4515. [Online]. Available:
  \url{http://proceedings.mlr.press/v97/mendis19a.html}
\BIBentrySTDinterwordspacing

\bibitem{Blanquer00}
\BIBentryALTinterwordspacing
J.~M. Blanquer and R.~C. Chalmers, ``{MOB}: A memory organization benchmark,''
  University of California, Santa Barbara, Tech. Rep., 2000. [Online].
  Available:
  \url{http://www.gnu-darwin.org/www001/ports-1.5a-CURRENT/devel/mob/work/mob-0.1.0/doc/mob.ps}
\BIBentrySTDinterwordspacing

\bibitem{John07}
\BIBentryALTinterwordspacing
T.~John and R.~Baumgartl, ``Exact cache characterization by experimental
  parameter extraction,'' in \emph{RTNS}, Nancy, France, 2007, pp. 65--74.
  [Online]. Available:
  \url{https://hal.inria.fr/inria-00168530/file/actes.pdf\#page=66}
\BIBentrySTDinterwordspacing

\bibitem{Zhang14}
\BIBentryALTinterwordspacing
Y.~Zhang, N.~Guan, and W.~Yi, ``Understanding the dynamic caches on {Intel}
  processors: Methods and applications,'' in \emph{Proceedings of the 12th IEEE
  International Conference on Embedded and Ubiquitous Computing}, ser. EUC
  '14.\hskip 1em plus 0.5em minus 0.4em\relax USA: IEEE Computer Society, Aug.
  2014, pp. 58--64. [Online]. Available:
  \url{http://dx.doi.org/10.1109/EUC.2014.18}
\BIBentrySTDinterwordspacing

\end{thebibliography}

\end{document}